\documentclass{aa}
\usepackage{graphicx}
\usepackage{psfig}
\usepackage{epsfig}
\begin{document}
\title{UV(IUE) spectra of hot post-AGB candidates 
\thanks{Based on observations obtained with the International
Ultraviolet Explorer (IUE), retrieved from the  Multimission Archive at STScI}}
\author{G. Gauba 
\and M. Parthasarathy} 
\institute{Indian Institute of Astrophysics, Koramangala, Bangalore 560034, 
India}
\offprints{G. Gauba,
\email{gsarkar@iucaa.ernet.in}}
\date{Received / Accepted}
\authorrunning{G. Gauba and M. Parthasarathy}
\titlerunning{IUE spectra of hot post-AGB candidates}

\abstract{Analysis of the low resolution UV(IUE) spectra of 15 hot post-AGB 
candidates is presented.
The UV(IUE) spectra of 10 stars suggest partial obscuration of
the hot stars due to circumstellar dust. The reddened 
continua of these 10 stars were used to model and estimate the 
circumstellar extinction. The circumstellar extinction law was found 
to be linear in $\lambda^{-1}$ in the case of 
IRAS13266-5551 (CPD-55 5588), IRAS14331-6435 (Hen3-1013),
IRAS16206-5956 (SAO 243756), IRAS17074-1845 (Hen3-1347),
IRAS17311-4924 (Hen3-1428), IRAS18023-3409 (LSS 4634), IRAS18062+2410 (SAO 85766),
IRAS18371-3159 (LSE 63), IRAS22023+5249 (LSIII +5224) and
IRAS22495+5134 (LSIII +5142). There seems to be no significant circumstellar
extinction in the case of IRAS17203-1534, IRAS17460-3114 (SAO 209306) and 
IRAS18379-1707 (LSS 5112). The UV(IUE) spectrum of IRAS12584-4837 (Hen3-847)
shows several emission lines including that of HeII. 
It may be a massive young OB-supergiant or a low mass star
in the post-AGB phase of evolution.  IRAS16206-5956 (SAO 243756) and 
IRAS 18062+2410 (SAO 85766)
show variability in the UV which in addition to stellar pulsations 
may be attributed to a dusty torus in motion around the hot central 
stars. The UV spectrum of the bipolar PPN, 
IRAS17423-1755 (Hen3-1475) indicates that the central B-type star 
is obscured by a dusty disk. 
The stars were placed on the log g$-$log T$_{\rm eff}$ diagram 
showing the post-AGB evolutionary tracks of Sch\"onberner. 
Terminal wind velocities of the stars were estimated from the CIV and NV 
stellar wind features. The presence of stellar wind in some of
these stars indicates ongoing mass-loss. 
}
\maketitle

\newpage
\section{Introduction}

Low and intermediate mass stars (M $\simeq$ 
0.8 - 8 M$_{\odot}$) pass through the post-asymptotic
giant branch (post-AGB) phase of evolution on their way to becoming 
planetary nebulae (PNe). From an analysis of the Infrared Astronomical 
Satellite Point Source Catalog (IRAS PSC) cooler post-AGB stars having G,F,A 
supergiant like character were first identified (Parthasarathy
\& Pottasch 1986, Lamers et al. 1986, Pottasch \& Parthasarathy 1988a,
Hrivnak et al. 1989). These stars were  found to have circumstellar 
dust shells with far-IR colors and flux distributions similar
to the dust shells of PNe. Later, from an 
analysis of IRAS data, Parthasarathy \& Pottasch (1989) found a
few hot (OB spectral types) post-AGB candidates. 
Their supergiant like character, the presence of cold detached
dust shells, far-IR colors similar to PNe and high galactic latitudes suggested 
that they may be in a post-AGB phase of evolution. Thus, there seems 
to be an evolutionary sequence ranging from the cooler 
G,F,A supergiant-like stars to hotter O-B types, evolving from the 
tip of the AGB towards young PN stage (Parthasarathy, 1993c). 

Pottasch et al. (1988b) and van der Veen \& Habing (1988) identified
a region of the IRAS color-color diagram (F(12$\mu$)/F(25$\mu$) $<$ 0.35
and F(25$\mu$)/F(60$\mu$) $>$ 0.3) which was mainly populated
by stars in transition from the AGB to the PN phase. Based on their
far-IR colors and low resolution optical spectra, several hot post-AGB 
candidates were identified (Parthasarathy \& Pottasch 1989, 
Parthasarathy 1993a, 1993c, Parthasarathy et al., 2000a).
The optical spectra of these objects show strong Balmer emission lines and 
in some cases low excitation nebular emission lines such as [NII] and [SII]
superposed on the OB stellar continuum. The absence of [OIII] 5007\AA~
line and the presence of low excitation nebular emission lines indicate
that photoionisation has just started. It is important to study these
stars in the UV to obtain better estimates of their temperatures and
to look for signatures of circumstellar reddening, mass-loss and 
stellar winds. The UV(IUE) spectra of 
some hot post-AGB stars (eg. Hen3-1357, Parthasarathy et al. 1993b, 1995,
Feibelman, 1995) have revealed violet shifted stellar wind P-Cygni profiles 
of CIV, SiIV and NV, indicating hot and fast stellar wind, 
post-AGB mass-loss and rapid evolution. In this paper we have analysed the UV(IUE) spectra 
of 15 hot post-AGB candidates. 

\section{Target Selection and Observations}

The hot post-AGB candidates in this paper (Table 1), were  
identified on the basis of their IRAS colors
(F(12$\mu$)/F(25$\mu$) $<$ 0.35 and F(25$\mu$)/F(60$\mu$) $>$ 0.3),
high galactic latitudes and OB-giant or supergiant 
spectra in the optical (Parthasarathy et al., 2000a)
with Balmer lines in emission.
Young massive OB supergiants are not expected at high galactic latitudes and
also they are not expected to have detached cold circumstellar dust shells.
High galactic latitude OB supergiants with detached dust shells
and far-IR colors similar to PNe were found to be in the
post-AGB phase of evolution (Parthasarathy, 1993c, 
Parthasarathy et al., 2000a). 

\begin{table*}
\begin{center}
\caption{Hot post-AGB candidates}
\begin{tabular}{|c|c|c|c|c|c|c|c|c|}
\hline 

Star No. & IRAS & Name & l & b & \multicolumn{4}{c|}{IRAS Fluxes (Jy.)} \\ \cline{6-9} 
         &     &      &   &   & 12$\mu$ & 25$\mu$ & 60$\mu$ & 100$\mu$ \\
\hline \hline

1. & 12584-4837 & Hen3-847     & 304.60&  +13.95& 36.07 & 48.75 & 13.04 & 3.31 \\
2. & 13266-5551 & CPD-55 5588   & 308.30&  +6.36 & 0.76  & 35.90 & 35.43 & 11.66\\
3. & 14331-6435 & Hen3-1013    & 313.89&  -4.20 & 4.04  & 108.70& 70.71 & 20.61\\
4. & 16206-5956 & SAO 243756   & 326.77&  -7.49 & 0.36L & 11.04 & 12.30 & 4.83\\ 
5. & 17074-1845 & Hen3-1347    &   4.10&  +12.26& 0.50  & 12.20 &  5.66 & 3.47:\\
6. & 17203-1534 &              & 8.55  &  +11.49& 0.32  & 10.70 & 6.88  & 3.37\\    
7. & 17311-4924 & Hen3-1428    & 341.41&  -9.04 & 18.34 & 150.70& 58.74 & 17.78\\ 
8. & 17423-1755 & Hen3-1475    &   9.36&  +5.78 &  7.05 &  28.31& 63.68 & 33.43\\ 
9. & 17460-3114 &  SAO 209306  & 358.42&  -1.88 & 6.26  & 20.82 & 12.20 &220.40L\\
10.& 18023-3409 & LSS 4634     & 357.61&  -6.31 & 0.26L & 2.94  & 1.82  & 25.64L\\ 
11.& 18062+2410 & SAO 85766    &  50.67&  +19.79& 3.98  & 19.62 & 2.90 & 1.00L \\  
12.& 18371-3159 & LSE 63       &  2.92 & -11.82 & 0.25L &  6.31 &  5.16 & 1.95\\
13.& 18379-1707 & LSS 5112     & 16.50 & -5.42  & 1.67  & 23.76 &  7.12 & 3.66L\\
14.& 22023+5249 & LSIII +5224  & 99.30 &  -1.96 & 1.02  & 24.69 & 14.52 & 3.93L\\
15.& 22495+5134 & LSIII +5142  & 104.84&  -6.77 & 0.54  & 12.37 &  7.18 & 3.12\\ 

\hline
\end{tabular}

\vspace{0.2cm}

\noindent \parbox{16cm}{A colon {\bf :} indicates moderate quality IRAS flux,
{\bf L} is for an upper limit}
\end{center}
\end{table*}

Low resolution ($\sim$ 6 $-$ 7\AA~), large aperture, UV(IUE) spectra of 
the hot post-AGB candidates from 1150\AA~ to 3200\AA~ were extracted 
from the Multimission Archive at STScI (Table 2). The spectra 
obtained by centering the stars in the 10\arcsec X 23\arcsec 
aperture were  processed using the IUE NEWSIPS (new spectral 
image processing system) pipeline which applies the 
signal weighted extraction technique (SWET) as well as the latest 
flux calibration and close-out camera sensitivity corrections.
An increased signal-to-noise (S/N) ratio of 10\% $-$ 50\%
has been demonstrated for low dispersion IUE spectra reprocessed 
with the NEWSIPS software (Nichols \& Linsky, 1996). Well exposed
IUE NEWSIPS spectra have S/N of $\sim$ 50 while weak, high-background, 
under-exposed spectra have S/N of $\sim$ 20 
(Nichols et al., 1994, Nichols \& Linsky, 1996). From our sample,
IRAS14331-6435 (Hen3-1013), IRAS17074-1845 (Hen3-1347), IRAS17203-1534, 
IRAS18023-3409 (LSS 4634), IRAS18379-1707 (LSS 5112) and 
IRAS22023+5249 (LSIII +5224) have S/N $\sim$ 20. IRAS12584-4837 (Hen3-847),
IRAS17460-3114 (SAO 209306) and IRAS18371-3159 (LSE 63) have well 
exposed spectra with S/N $\sim$ 50. The spectra of the remaining hot 
post-AGB candidates are of intermediate quality with S/N $\sim$ 30.
The LWP spectra 18412 and 27936 of IRAS12584-4837 and IRAS18371-3159 
respectively, were saturated and have not been used in the analysis. 
Line-by-line images were inspected for spurious features. 

\begin{table*}
\begin{center}
\caption{Log of observations}
\begin{tabular}{|c|c|c|c|c|c|} \hline
IRAS & Camera & Image & Aperture & DateObs & Exposure Time(s) \\ 
\hline \hline

12584-4837 & SWP & 39271 & LARGE & 21 July 1990 & 2999.781\\
= Hen3-847 & LWP & 18412 & LARGE & 21 July 1990 & 1799.659\\
           & LWP & 19726 & LARGE & 10 Feb 1991 &  299.704\\

\hline

13266-5551 & SWP & 39270 & LARGE & 20 July 1990 & 1799.652\\
= CPD-55 5588& LWP & 18411 & LARGE & 20 July 1990 & 1199.595\\

\hline

14331-6435 & SWP & 33601 & LARGE & 22 May 1988 & 2399.717\\
= Hen3-1013 & LWP & 13295 & LARGE & 22 May 1988 & 1199.595\\

\hline

16206-5956 & SWP & 33953 & LARGE & 21 July 1988 & 2399.717\\
= SAO 243756 & SWP & 50195 & LARGE & 12 March 1994  & 2399.716\\
           & SWP & 50640 & LARGE & 28 April 1994  & 3599.844\\
           & LWP & 13714 & LARGE & 21 July 1988   & 1199.595\\
           & LWP & 26175 & LARGE & 19 August 1993 & 1019.781\\
           & LWP & 27667 & LARGE & 12 March 1994 & 599.531\\

\hline

17074-1845 & SWP & 39276 & LARGE & 21 July 1990 & 1799.652\\
= Hen3-1347 & LWP & 18418 & LARGE & 21 July 1990 & 599.531\\

\hline

17203-1534 & SWP & 50657 & LARGE & 30 April 1994 & 2399.716\\
           & LWP & 28020 & LARGE & 30 April 1994 & 1199.595\\

\hline
 
17311-4924 & SWP & 33600 & LARGE & 21 May 1988 & 2099.480\\
= Hen3-1428 & SWP & 48127 & LARGE & 17 July 1993 & 3599.844\\
           & LWP & 13294 & LARGE & 22 May 1988 & 959.570\\
           & LWP & 25934 & LARGE & 17 July 1993 & 899.768\\

\hline

17423-1755 & SWP & 35860 & LARGE & 26 March 1989 & 2099.480\\
= Hen3-1475&     &       &       &               &         \\ 

\hline

17460-3114 & SWP & 54660 & LARGE & 12 May 1995 & 89.572\\
= SAO 209306 &   &       &       &             &       \\

\hline

18023-3409 & SWP & 55458 & LARGE & 9 August 1995 & 4199.499\\
= LSS 4634 & LWP & 31272 & LARGE & 9 August 1995 & 1559.634\\

\hline

18062+2410 & SWP & 44446 & LARGE & 21 April 1992 & 1200 \\
= SAO 85766 & SWP & 55916 & LARGE & 12 Sept. 1995 & 1800 \\
           & LWP & 22865 & LARGE & 21 April 1992 &  300 \\
           & LWP & 31454 & LARGE & 12 Sept. 1995 &  600 \\

\hline

18371-3159 & SWP & 50588 & LARGE & 19 April 1994 & 4799.563\\
= LSE 63   & LWP & 27936 & LARGE & 19 April 1994 & 2399.723\\
           & LWP & 27972 & LARGE & 23 April 1994 & 899.768\\

\hline

18379-1707 & SWP & 47531 & LARGE & 23 April 1993 & 3899.672\\
= LSS 5112 & LWP & 25398 & LARGE & 23 April 1993 & 1199.595\\

\hline

%
%
%
22023+5249 & SWP & 48454 & LARGE & 24 August 1993 & 2399.716\\
= LSIII +5224 & SWP & 48593 & LARGE & 9 Sept 1993 & 4799.563\\
           & SWP & 55915 & LARGE & 12 Sept 1995 & 7199.819\\
           & LWP & 26209 & LARGE & 24 August 1993  & 1199.595\\
           & LWP & 31453 & LARGE & 12 Sept 1995 & 2399.723\\

\hline

22495+5134 & SWP & 48453 & LARGE & 24 August 1993 & 2399.716\\
= LSIII +5142 & LWP & 26322 & LARGE & 9 Sept. 1993   & 2099.487\\

\hline

\end{tabular}
\end{center}
\end{table*}

\section{Analysis}

Tables 3, 4a and 4b list the V magnitudes and B$-$V values of the 15 hot 
post-AGB candidates from literature. The optical 
spectral types are mainly from Parthasarathy et al. (2000a). 
For the optical spectral types of the stars, the 
corresponding intrinsic B$-$V values, (B$-$V)$_{\rm o}$, were taken from 
Schmidt-Kaler (1982). Using the observed and intrinsic B$-$V values 
we derived the total (interstellar plus circumstellar) extinction, 
E(B$-$V)$_{\rm total}$ (=(B$-$V)$_{\rm obs}$ $-$ (B$-$V)$_{\rm o}$) 
towards these stars. These values were compared with 
interstellar extinction (E(B$-$V)$_{\rm I.S.}$) at the galactic latitudes and
longitudes of these stars, estimated using the Diffuse Infrared 
Background Experiment (DIRBE)/IRAS dust maps (Schlegel et al., 1998). 
The DIRBE/IRAS dust maps do not give reliable estimates of the 
interstellar extinction for $|$b$|$ $<$ 5$^{\circ}$. The accuracy
of the DIRBE/IRAS extinction estimates are 16\%.
Comparing E(B$-$V)$_{\rm total}$ and E(B$-$V)$_{\rm I.S.}$
we found considerable circumstellar extinction in most cases 
(Tables 4a and b).

The 2200\AA~ feature in the UV gives an estimate of the interstellar
extinction. The merged LWP and SWP spectra of the hot post-AGB candidates
were dereddened (Tables 3, 4a and b) using the 2200\AA~ feature in the UV.
Using the UNRED routine in the IUE data analysis software package, 
we adopted the average extinction law by
Seaton (1979) and tried different values of E(B$-$V) till the
2200\AA~ bump in the UV disappeared and smooth continua from 1150\AA~
to 3200\AA~ were obtained for all the stars.
In the absence of an LWP spectrum of IRAS17460-3114 (SAO 209306), 
the SWP spectrum of the star was dereddened using E(B$-$V)$_{\rm total}$
((B$-$V)$_{\rm obs}-$(B$-$V)$_{\rm o}$)=0.54. 
In the rest of this paper, "dereddened spectra" would always refer
to the observed IUE spectra corrected for interstellar extinction
as determined from the 2200\AA~ feature in the UV.

For stars with multiple LWP and SWP spectra, we overplotted the
observed LWP (and SWP) spectra of each star to check for variability.
The LWP and SWP spectra of IRAS16206-5956 (SAO 243756)
and IRAS18062+2410 (SAO 85766) were found to show variation (Fig.1). 
Coadded LWP and SWP spectra of IRAS17311-4924 and IRAS22023+5249 were used
in the analysis.

\begin{figure*}
\epsfig{figure=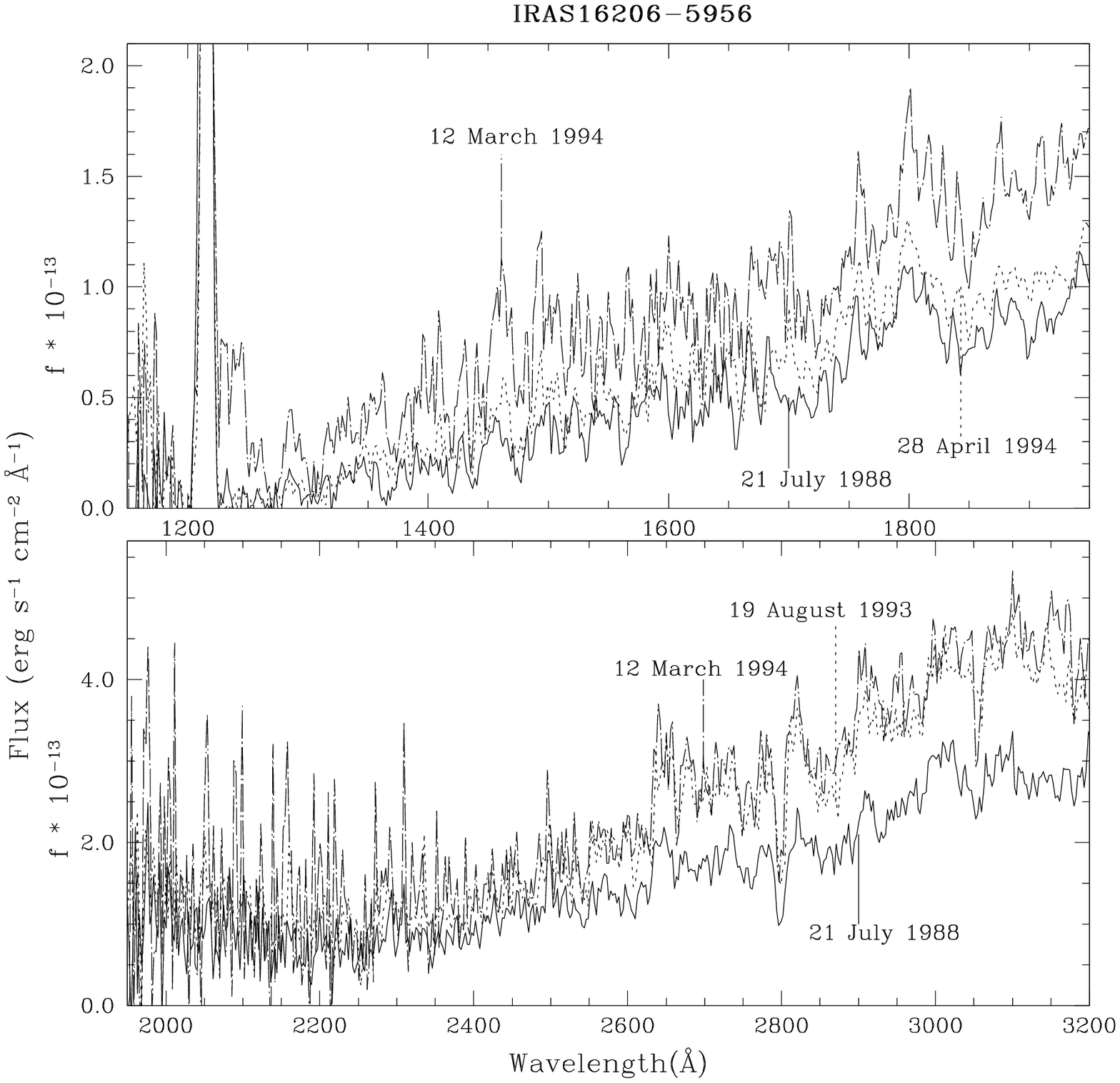,width=15cm,height=10cm}

\vspace{0.3cm}

\epsfig{figure=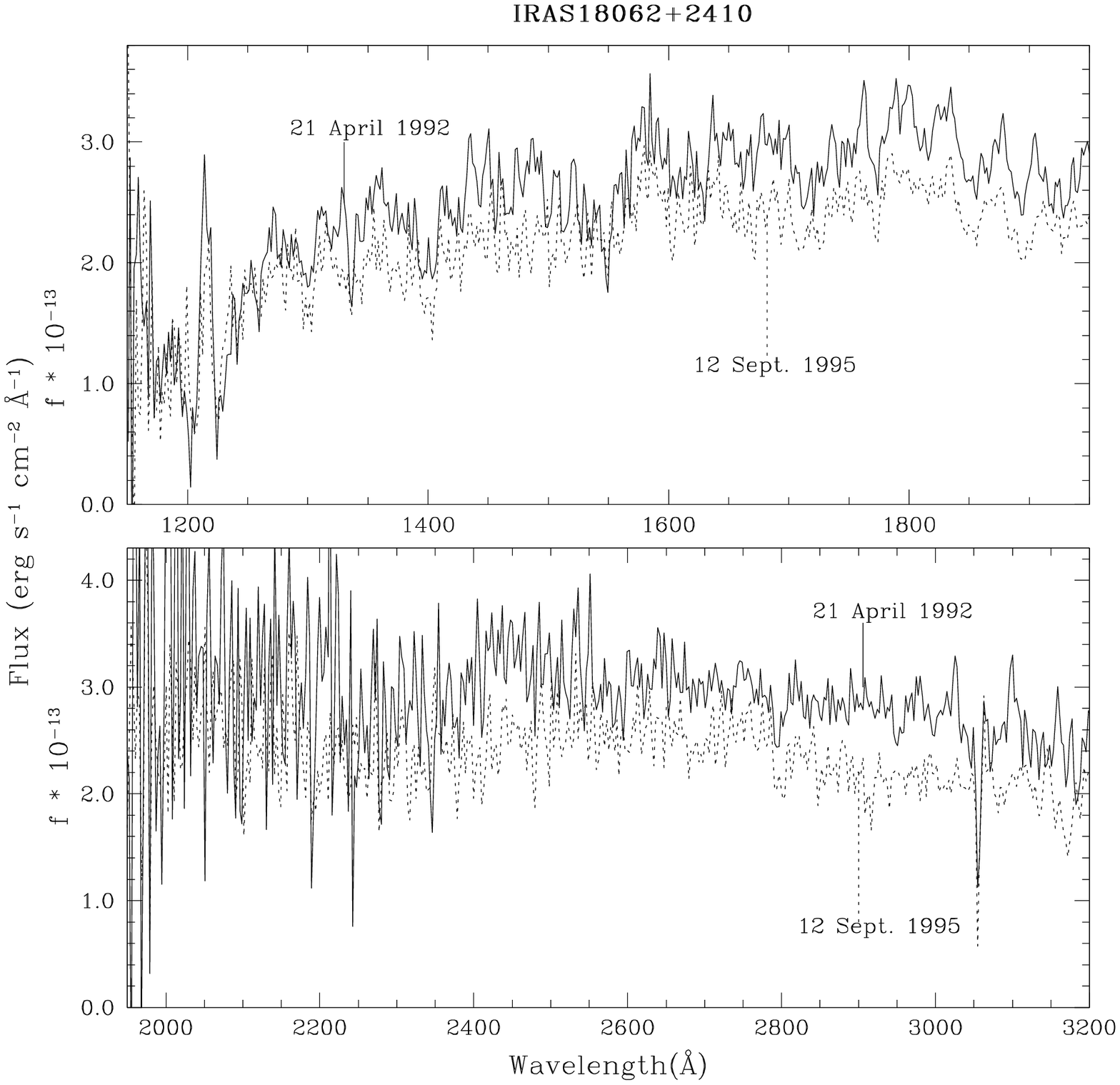,width=15cm,height=10cm}
\caption{The observed LWP and SWP spectra of IRAS16206-5956(SAO 243756)
and IRAS18062+2410(SAO 85766) showing the variability.}
\end{figure*}

The dereddened merged spectra of the hot post-AGB candidates were compared 
with the dereddened spectra of standard stars from the atlas by Heck et al. (1984) 
The spectral types of the standard stars chosen for comparison were as 
close as possible to the optical spectral types of the hot post-AGB candidates. 
Tables 3, 4a and 4b list the standard stars used for comparison, their spectral 
types and E(B$-$V) values. For comparison, the flux of the standard 
stars were scaled in accordance with the difference between the V magnitudes 
(after correcting for interstellar exintinction as determined from the 2200\AA~
feature) of the standard star and the corresponding hot post-AGB candidate.

\subsection{Spectral features in the UV}

The spectra from 1150\AA~ to 3200\AA~ dereddened with E(B$-$V) 
determined from the 2200\AA~ feature are shown in Figs. 2, 3, 5a and 6a. 
Lines of NV(1240\AA~), CII(1335\AA~), SiIV(1394, 1403\AA~), CIV(1550\AA~), 
NIV(1718\AA~), FeII(2586-2631\AA~) and MgII(2800\AA~) typical of hot stars 
(Heck et al., 1984) and central stars of PNe were identified in the spectra of 
these stars. The UV line strengths in the hot post-AGB candidates were compared 
with the line strengths in the standard stars. The absence of CIV in the B3-supergiant
star, IRAS14331-6435 (Hen3-1013) alongwith near normal line strengths of 
SiIV and NIV suggests the underabundance of carbon in this star. From the
SiIV and CIV features it appears that IRAS17074-1845 (Hen3-1347) and 
IRAS17460-3114 (SAO 209306) may be slightly metal deficient. Based on its 
optical spectra, IRAS18062+2410 (SAO 85766) was found to be metal poor and
underabundant in carbon (Parthasarathy et al. 2000b, Mooney et al., 2002). 
In the UV also the line strengths appear to be weaker compared to a standard 
B1-supergiant. Metal deficiency has been observed in some high latitude 
hot post-AGB stars (see eg. McCausland et al. 1992, Napiwotzki et al. 1994). 

\begin{figure*}
\epsfig{figure=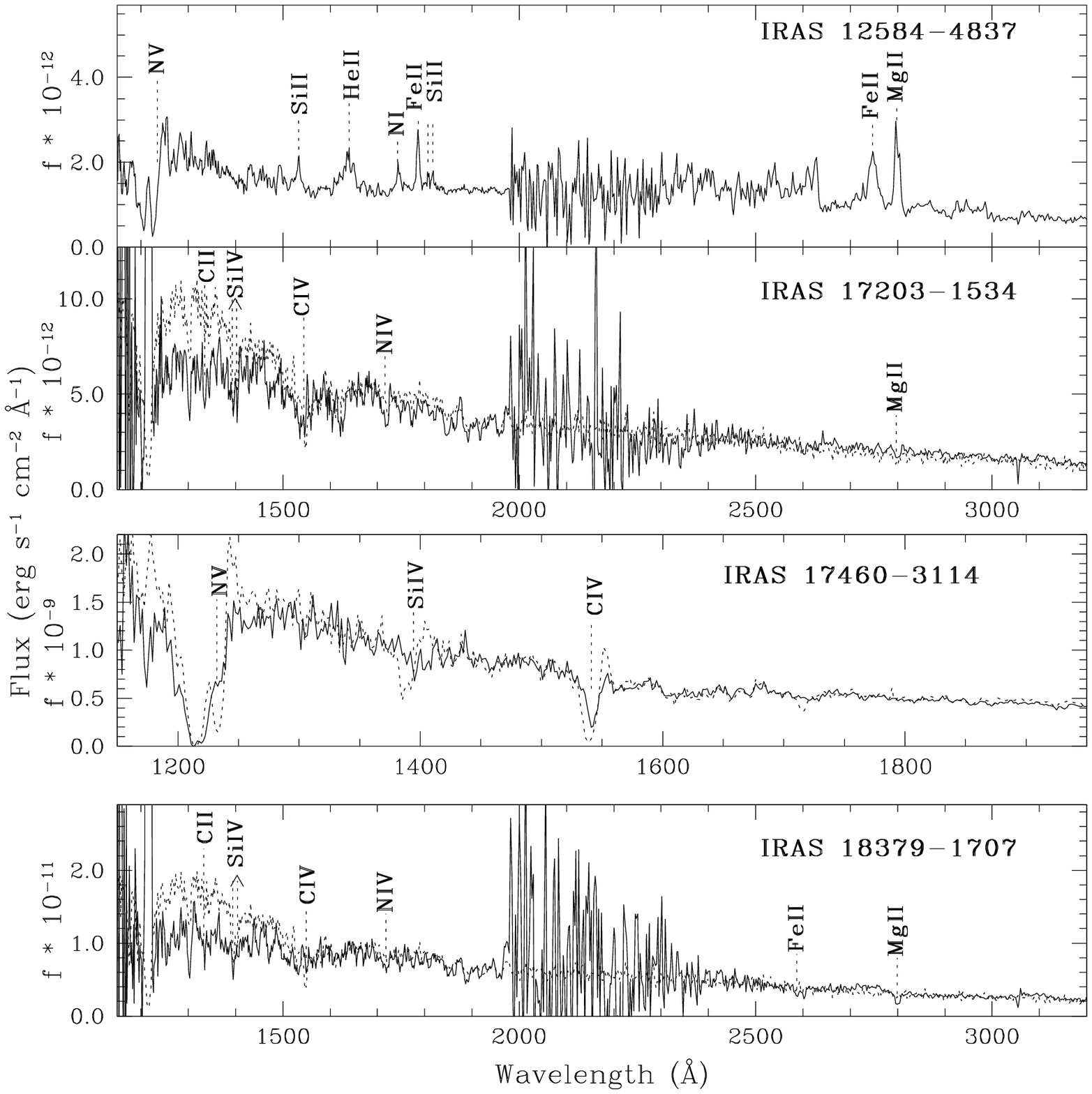, width=17cm, height=20cm}
\caption{The dereddened UV(IUE) spectra of hot post-AGB candidates 
with negligible circumstellar extinction (solid lines) compared with the
dereddened UV(IUE) spectra of standard stars (dotted lines) from the atlas 
by Heck et al. (1984, Table 3). The spectra were dereddened using
E(B$-$V) estimated from the 2200\AA~ feature in the UV. IRAS17460-3114
was dereddened using E(B$-$V)$_{\rm total}$ for the star. IRAS12584-4837
is a Be star in the optical.}
\end{figure*}

\begin{figure*}
\epsfig{figure=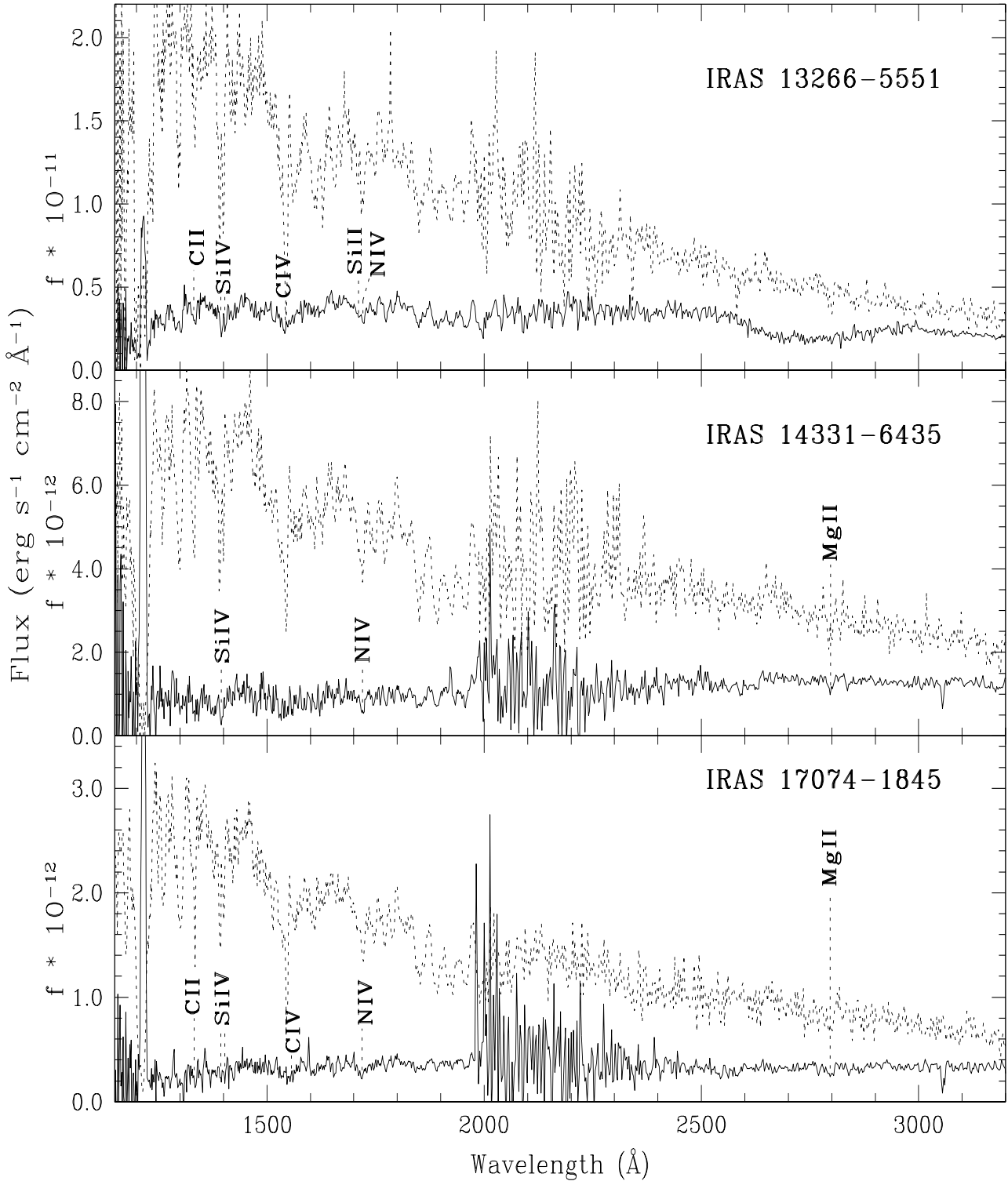, width=17cm, height=20cm}
\caption{The dereddened UV(IUE) spectra of hot post-AGB candidates
(solid line) plotted alongwith the dereddened spectra of standard
stars (dotted line) of similar optical spectral types from the 
atlas by Heck et al. (1984, Table 4a). The spectra have been
dereddened using the 2200\AA~ feature in the UV. Notice the reddened 
continua of the hot post-AGB candidates in comparison with the standard
stars. The hot central star of IRAS17423-1755 is not observed
in the UV, possibly due to obscuration of the star by a dusty torus}.
\end{figure*}

\setcounter{figure}{2}
\begin{figure*}
\renewcommand{\thefigure}{\arabic{figure}}
\epsfig{figure=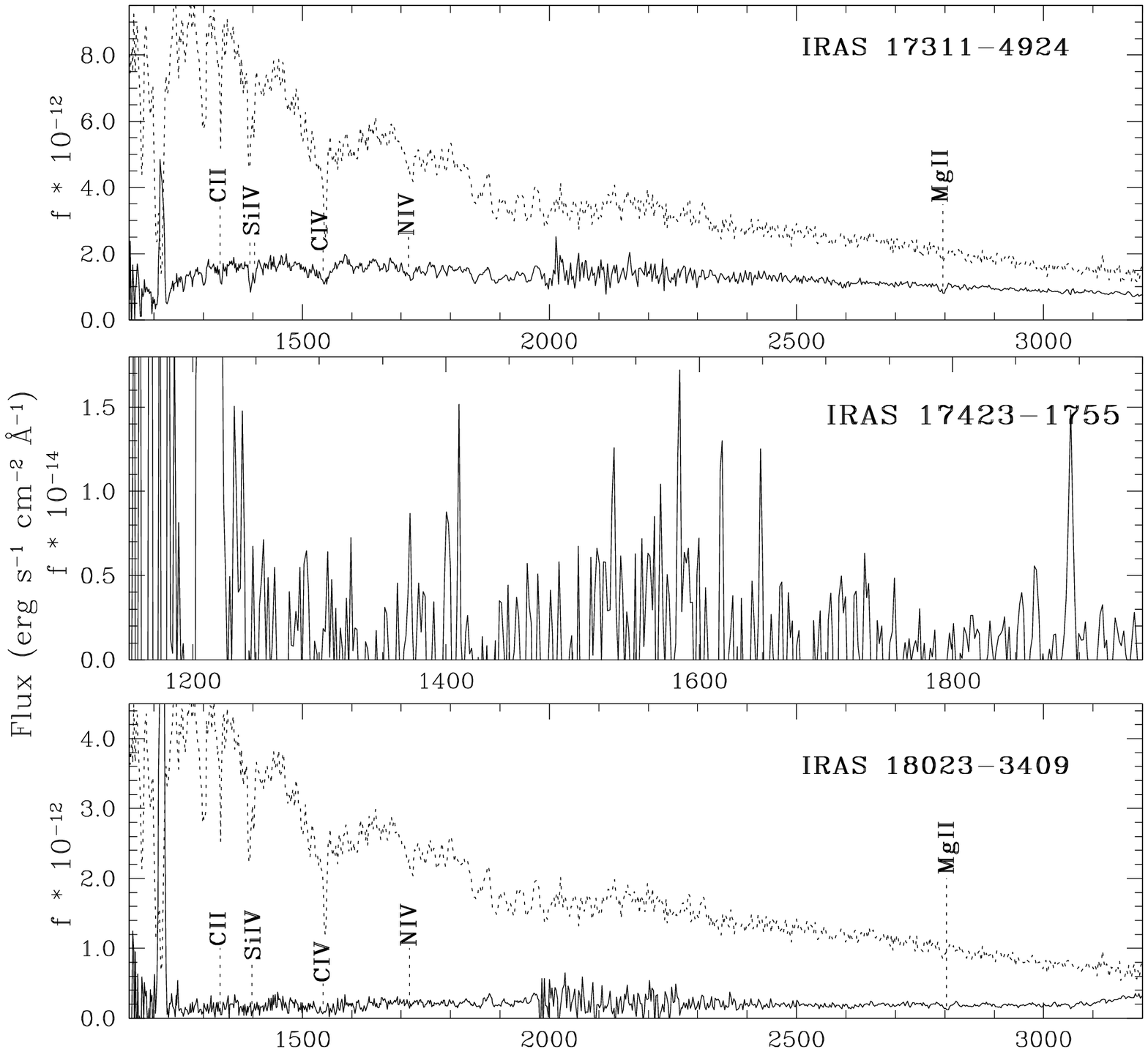, width=17cm, height=20cm}
\caption{contd....}
\end{figure*}

\setcounter{figure}{2}
\begin{figure*}
\renewcommand{\thefigure}{\arabic{figure}}
\epsfig{figure=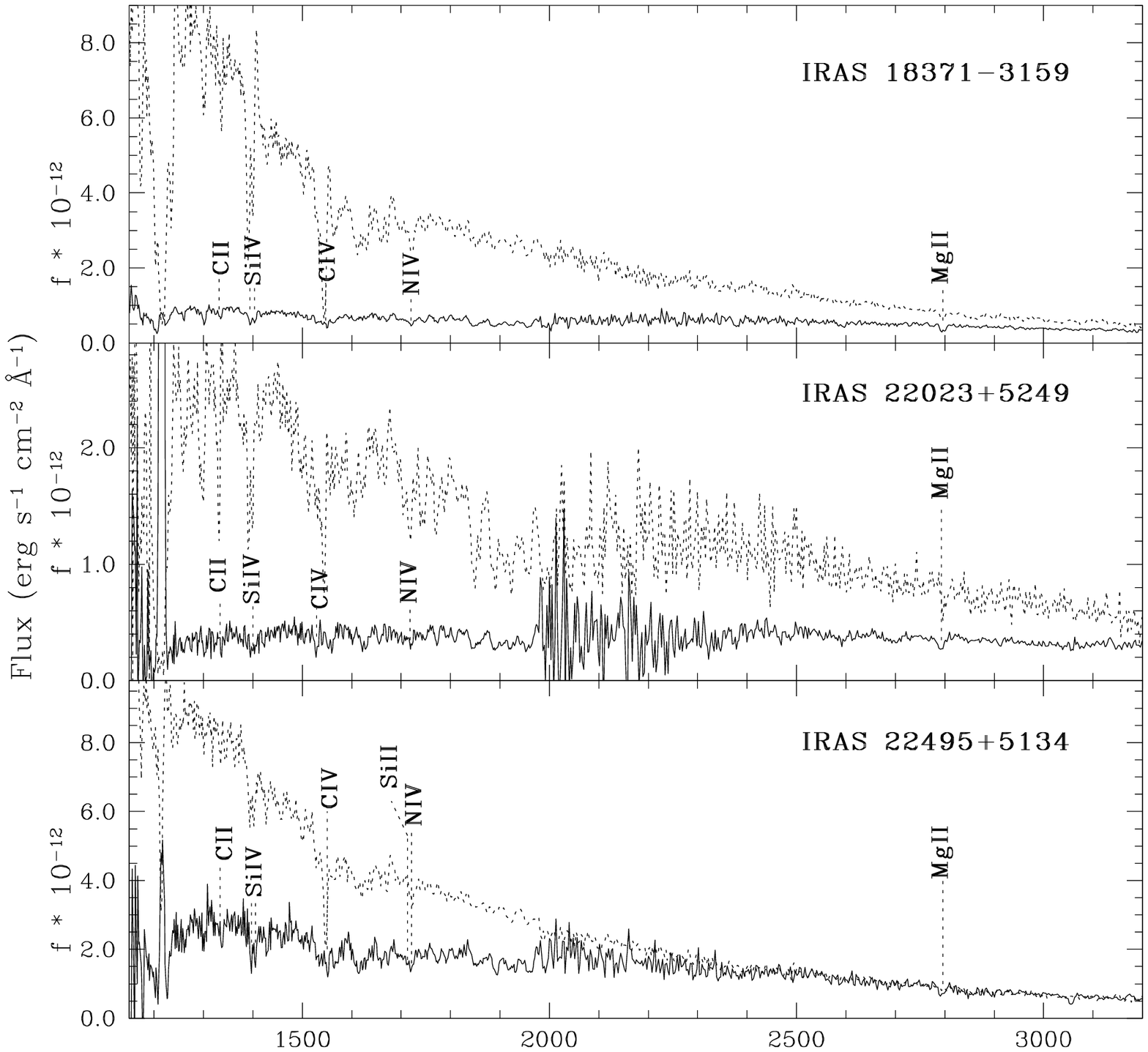,width=17cm,height=20cm}
\caption{contd....}
\end{figure*}

\subsection{Stars with negligible circumstellar extinction}

The UV continua and spectral features of IRAS17203-1534, 
IRAS17460-3114 (SAO 209306) and IRAS18379-1707 (LSS 5112) were in good agreement with 
the dereddened UV(IUE) spectra of standard stars (Table 3) of similar 
optical spectral types (Fig. 2). The E(B$-$V) values of these stars 
determined from the 2200\AA~ feature are nearly the same as E(B$-$V)$_{\rm total}$
(Table 3) suggesting negligible extinction of starlight due to 
circumstellar dust in these three cases. 
Emission lines of SiII(1533, 1808, 1817\AA~), HeII(1640\AA~), NI(1743\AA~) and 
FeII(1785, 2746\AA~) in the spectrum of IRAS12584-4837 (Hen3-847) indicate the presence of hot
plasma or a nebula.

\begin{table*}
\begin{center}
\caption{Hot post-AGB candidates with negligible circumstellar extinction}
\begin{tabular}{|c|c|c|c|c|c|c|c|c|c|} 
\hline
IRAS & V & (B$-$V)$_{\rm obs}$ & E(B$-$V) & E(B$-$V) & Sp. Type & E(B$-$V) & Standard & Sp.Type & E(B$-$V) \\  
     &mag&                 &  I.S.    &(2200\AA~)& Optical  & total    & star     & std. star& std. star\\ 
\hline \hline
12584-4837 & 10.58$^{a}$ & 0.07$^{a}$&0.18& 0.25 & Be$^{1}$& --   & --       & --    & --   \\
17203-1534 & 12.02$^{a}$& 0.35$^{a}$& 0.44& 0.56 & B1IIIpe & 0.61 & HD173502 & B1II  & 0.09 \\
17460-3114 &  7.94$^{b}$& 0.23$^{b}$& --  & 0.54 & O8III   & 0.54 & HD162978 & O7.5II& 0.35 \\  
18379-1707 & 11.93$^{b}$& 0.45$^{b}$& --  & 0.70 & B1IIIpe & 0.71 & HD173502 & B1II  & 0.11 \\

\hline
\end{tabular}
\noindent \parbox{17cm}
{Photometry is from : $^{a}$Hog et al.(2000); $^{b}$Reed(1998)
Spectra types of the hot post-AGB candidates are from Parthasarathy et al. (2000a)
except $^{1}$Kazarovets et al. (2000).} 

\end{center}
\end{table*}

\setcounter{table}{3}
\begin{table*}
\renewcommand{\thetable}{\arabic{table}a}
\begin{center}
\caption{Hot post-AGB candidates with observable circumstellar extinction}
\begin{tabular}{|c|c|c|c|c|c|c|c|c|c|c|} 
\hline
IRAS & V & (B$-$V)$_{\rm obs}$ & E(B$-$V) & E(B$-$V) & Sp. Type & E(B$-$V) & Standard & Sp.Type & E(B$-$V) & E(B$-$V)\\  
     &mag&                 &  I.S.    &(2200\AA~)& Optical  & total    & star     & std. star& std. star & C.S. \\ 
\hline \hline
13266-5551 & 10.68$^{b}$ & 0.31$^{b}$  & 0.53   & 0.38 & B1Ibe  & 0.51 & HD77581  & B0.5Ia & 0.77 & 0.15 \\
14331-6435 & 10.90$^{b}$ & 0.58$^{b}$  & --     & 0.44 & B3Ie   & 0.71 & HD198478 & B3Ia   & 0.58 & 0.26 \\
17074-1845 & 11.47$^{a}$ & 0.46$^{a}$  & 0.28   & 0.24 & B3IIIe & 0.66 & HD51309  & B3II   & 0.13 & 0.39 \\
17311-4924 & 10.68$^{c}$ & 0.40$^{c}$  & 0.22   & 0.28 & B1IIe  & 0.66 & HD51283  & B2III  & 0.02 & 0.39\\
17423-1755 & 12.64$^{d}$ & 0.66$^{d}$  & 0.67   & --   & Be     & --   &  --      & --     & --   & -- \\
18023-3409 & 11.55$^{b}$ & 0.46$^{b}$  & 0.44   & 0.30 & B2IIIe & 0.70 & HD51283  & B2III  & 0.02 & 0.43 \\ 
18371-3159 & 11.98$^{a}$ & 0.11$^{a}$  & 0.15   & 0.13 & B1Iabe & 0.30 & HD122879  & B0Ia   & 0.42 & 0.11 \\
22023+5249 & 12.52$^{a}$ & 0.69$^{a}$  & --     & 0.37 & B$^{1}$   & --   & HD41117  & B2Ia   & 0.52 & 0.44\\
22495+5134 & 11.78$^{a}$ & 0.22$^{a}$  & 0.357  & 0.33 & PN$^{2}$  & --   & HD38666  & 09V    & 0.02 & 0.20 \\

\hline
\end{tabular}
\noindent \parbox{17cm}
{Photometry is from : $^{a}$Hog et al.(2000); $^{b}$Reed(1998); $^{c}$Kozok (1985a); 
$^{d}$ Gauba et al.(2003)\\
Spectral types of the hot post-AGB candidates are from Parthasarathy et al. (2000a) except
$^{1}$Simbad database; $^{2}$Acker et al.(1992)}

\end{center}
\end{table*}

\setcounter{table}{3}
\begin{table*}
\renewcommand{\thetable}{\arabic{table}b}
\begin{center}
\caption{Hot post-AGB candidates showing variation in the UV}
\begin{tabular}{|c|c|c|c|c|c|c|c|c|c|c|} 
\hline

IRAS & V & (B$-$V)$_{\rm obs}$ & E(B$-$V) & E(B$-$V) & Sp. Type & E(B$-$V) & Standard & Sp.Type & E(B$-$V) & E(B$-$V)\\  
     & mag&                & I.S.     &(2200\AA~)& Optical  & total    & star     & std. star&std. star& C.S. \\ 
\hline \hline
16206-5956&9.76$^{a}$&0.31$^{a}$ & 0.22   & 0.13 & A0Ia$^{1}$& 0.29 & HD21389 & A0Ia & 0.79 & 0.54\\ 
          &          &           &        &      &           &      &          &      &      &  (21/07/88) \\ \cline{11-11} 
          &          &           &        &      &           &      &          &      &      & 0.24 \\ 
          &          &           &        &      &           &      &          &      &      & (12/03/94) \\
\hline
18062+2410&11.54$^{b}$&0.05$^{b}$ & 0.11   & 0.08 & B1I$^{2}$ & 0.24 & --      & --   & --   & 0.03 \\
          &           &           &        &      &           &      &         &      &      & (21/04/92) \\ \cline{11-11} 
          &           &           &        &      &           &      &         &      &      & 0.30\\
          &           &           &        &      &           &      &         &      &      & (12/09/95)\\
\hline
\end{tabular}
\noindent \parbox{17cm}
{Photometry is from : $^{a}$Reed(1998); $^{b}$Arkhipova et al.(1999)\\
Spectral types are from : $^{1}$Schild et al.(1983); $^{2}$Parthasarathy et al. (2000b)}

\end{center}
\end{table*}

\subsection{Stars with circumstellar extinction}

The UV(IUE) spectra of 10 stars (IRAS13266-5551 (CPD-55 5588), IRAS14331-6435 (Hen3-1013), 
IRAS16202-5956 (SAO 243756), IRAS17074-1845 (Hen3-1347), IRAS17311-4924 (Hen3-1428),
IRAS18023-3409 (LSS 4634), IRAS18062+2410 (SAO 85766), IRAS18371-3159 (LSE 63), 
IRAS22023+5249 (LSIII +5224), IRAS22495+5134 (LSIII +5142)),
dereddened using the 2200\AA~ feature in the UV,  
showed considerably reddened continua in comparison with the  
dereddened spectra of standard stars of similar optical spectral
types (Figs. 3, 5 and 6). Comparing the interstellar 
extinction estimates from the 2200\AA~ feature with the total 
extinction (E(B$-$V)$_{\rm total}$) towards these stars, we find that these
stars have considerable UV deficiency and circumstellar extinction.
The hot central star of the bipolar proto-planetary nebula (PPN), 
IRAS17423-1755 (Hen3-1475) was not detected 
in a 35 minute exposure with the SWP camera. This may be due to obscuration 
of the central star by a dusty disk. HST WFPC2 images of the object showed 
the presence of a dusty torus with a spatial extent of 2$\arcsec$ 
(Borkowski et al., 1997). 

\subsubsection{Modelling the circumstellar extinction}
 
To account for the observed UV deficiency in the 10 hot post-AGB candidates
mentioned in Sec. 3.3 and to understand the shape of the UV continuum in
these stars, we investigated the circumstellar extinction law in these cases.
Waters et al. (1989) modelled the circumstellar extinction in the case of
the post-AGB star, HR4049. We followed the same procedure here.

We plotted the logarithmic difference between the dereddened UV flux 
of a hot post-AGB candidate (normalised to its V-band flux) and the
dereddened UV flux of the corresponding standard star (normalised to its V-band flux),
i.e.\\ $\Delta = {\rm log (f_{\lambda}/f_{v})_{star} - log (f_{\lambda}/f_{v})_{standard}}$
Vs. $\lambda ^{-1}$. \\ Fig. 4 shows the plot of the logarithmic flux deficiency due 
to circumstellar dust from 3.2 to 8 $\mu^{-1}$ for the 10 stars. Best fit
lines were obtained by minimising the chi-square error statistic. Eg.,  
in the case of IRAS17311-4924 (Hen3-1428), we obtained, $\Delta$ = 0.07 $-$ 0.10$\lambda^{-1}$.
To derive the circumstellar extinction (E(B$-$V)$_{\rm C.S.}$) in magnitudes (Tables 4a and b)
$\Delta$ had to be multiplied by $-$2.5 and $\lambda$ = 0.44$\mu$ was used.

\begin{figure*}
\epsfig{figure=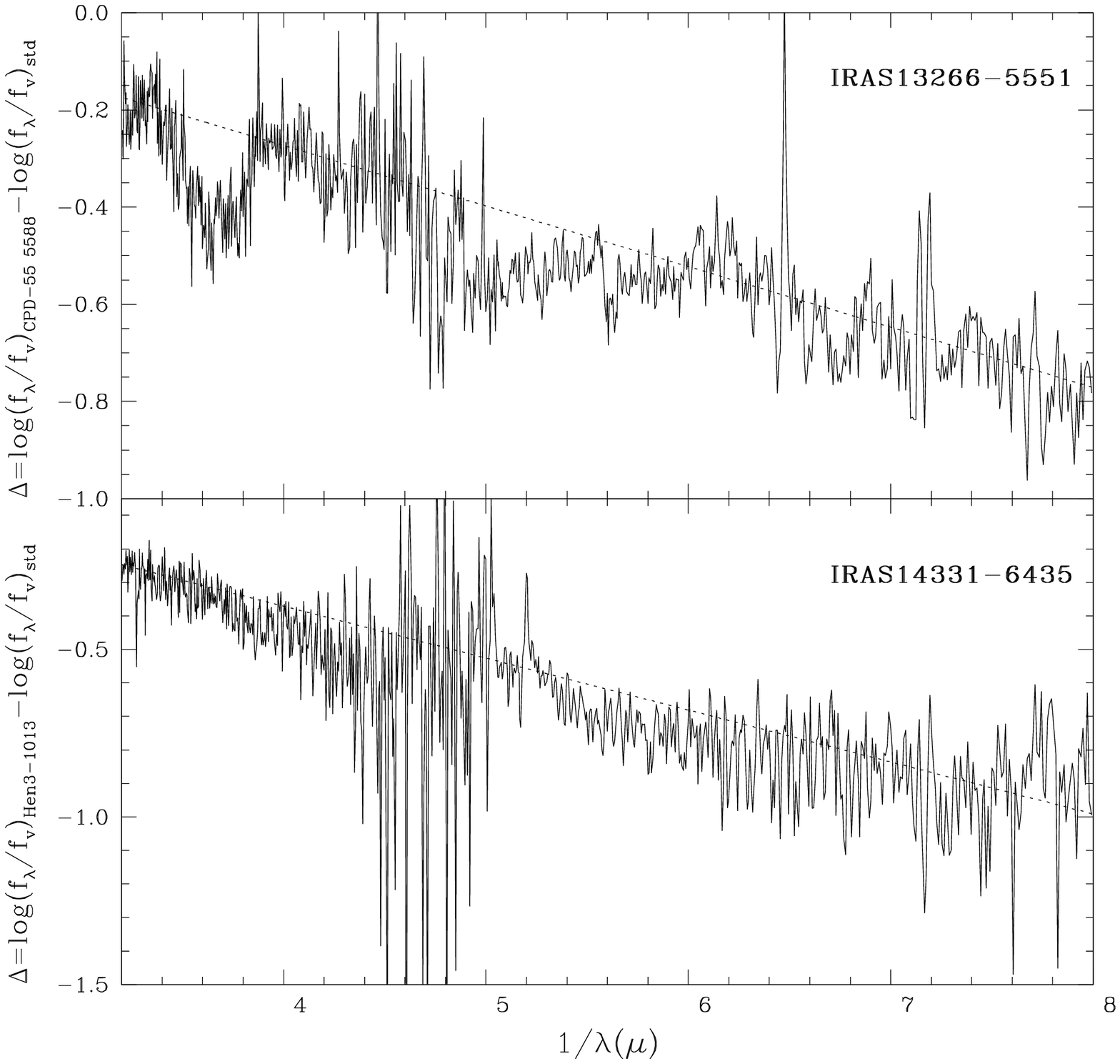,width=17cm,height=20cm}
\caption{10 hot post-AGB candidates showed a pronounced UV deficiency
when compared with standard stars of similar optical spectral types. 
Here, we have plotted the logarithmic flux deficiency of these stars in the
UV from 3.2 to 8 $\mu^{-1}$. The straight line fits are obtained by minimising
the chi-square statistic. We find that the flux deficiency in the UV is
proportional to $\lambda^{-1}$. The bump at $\sim$ 3.7$\mu^{-1}$ in the plot of
IRAS13266-5551 is due to the saturated LWP spectrum of the star. This region was
not used in obtaining a fit. Similarly, for IRAS18023-3409, the LWP spectrum
appears to have saturation effects. For IRAS18371-3159, the bump from
$\sim$ 5 to 5.6$\mu^{-1}$ is observed because of spurious broad absorption features 
from 1800\AA~ to 1950\AA~ in the SWP spectrum of the star.}
\end{figure*}

\setcounter{figure}{3}
\begin{figure*}
\renewcommand{\thefigure}{\arabic{figure}}
\epsfig{figure=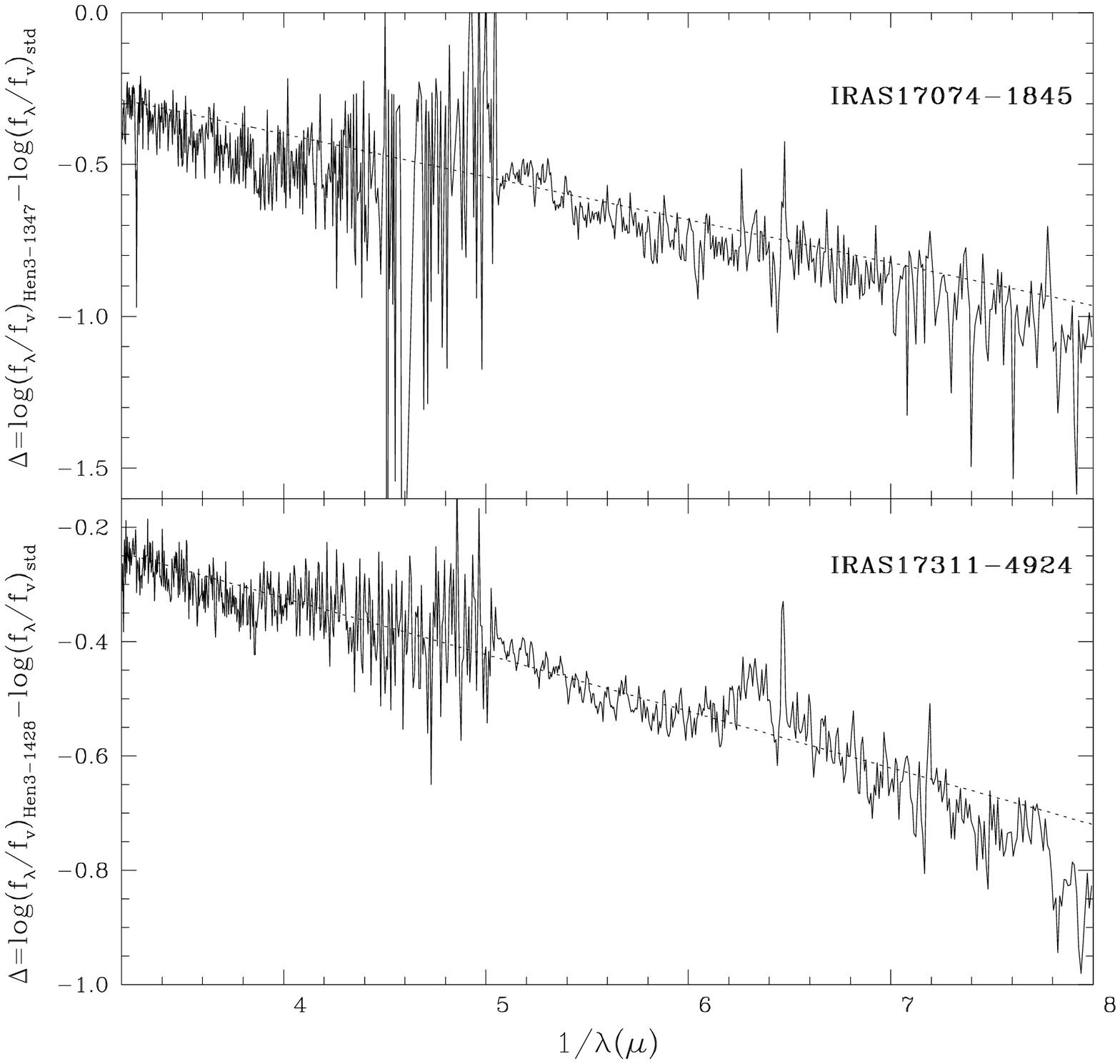,width=17cm,height=20cm}
\caption{contd...}
\end{figure*}

\setcounter{figure}{3}
\begin{figure*}
\renewcommand{\thefigure}{\arabic{figure}}
\epsfig{figure=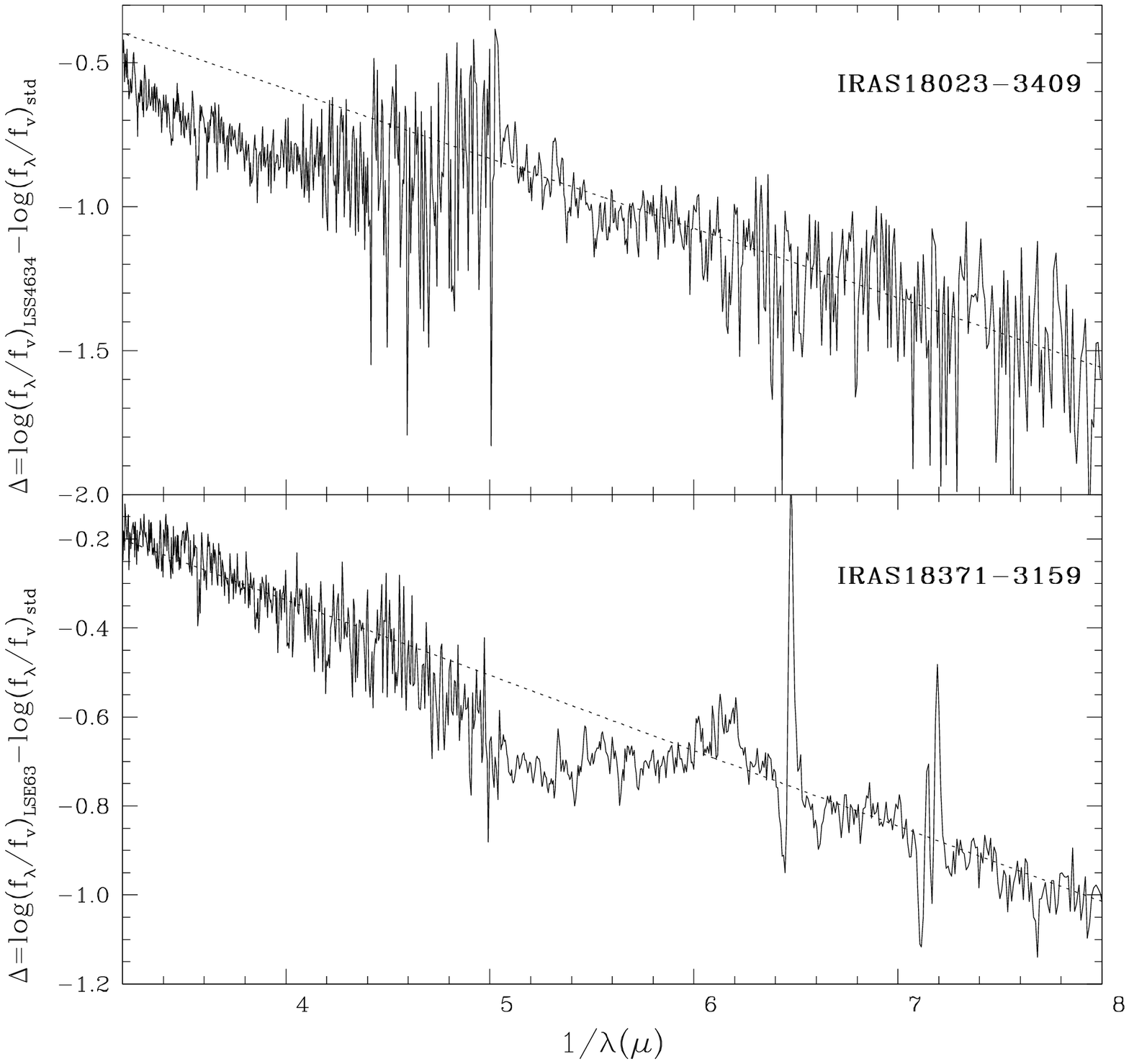,width=17cm,height=20cm}
\caption{contd...}
\end{figure*}

\setcounter{figure}{3}
\begin{figure*}
\renewcommand{\thefigure}{\arabic{figure}}
\epsfig{figure=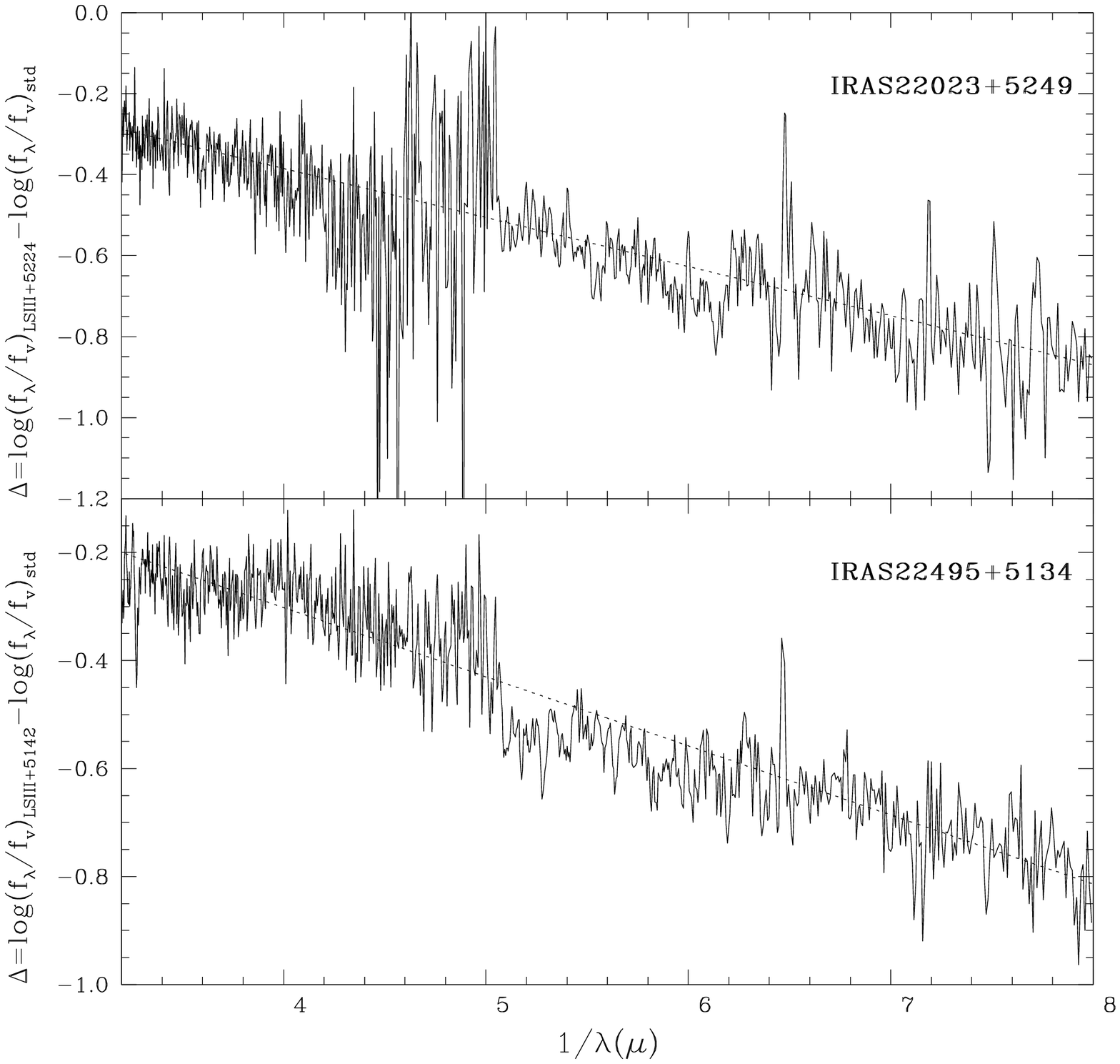,width=17cm,height=20cm}
\caption{contd...}
\end{figure*}

For IRAS22023+5249 (LSIII +5224), only a B spectral type is listed in literature. We 
compared the spectrum
of this star with that of a B2-supergiant standard star. IRAS22495+5134 (LSIII +5142) 
was detected as a PN with an angular extent of 0.5\arcsec (Tylenda \& 
Stasinska, 1994). Central stars of PNe have temperatures in excess of
$\sim$ 30000K corresponding to spectral types of O9 or hotter.
We compared the UV(IUE) spectra of this star with a standard O9V star (HD38666).

For each of the 10 hot post-AGB candidates we found that the 
circumstellar extinction varies as $\lambda^{-1}$ (Fig. 4).
The derived E(B$-$V)$_{\rm C.S.}$ values in Table 4a account well for the difference 
between the total and the interstellar extinction (from the 2200\AA~ feature) 
values. E(B$-$V)$_{\rm C.S.}$ values in Table 4b are in excess of the
difference between E(B$-$V)$_{\rm total}$ and the interstellar extinction
from 2200\AA~. This may be because of the variable nature of these stars and because
the V and (B$-$V) magnitudes at each epoch of the 
IUE observations are not known. Mean V and (B$-$V) magnitudes from literature 
have been used for each of these two stars (see Sec. 3.3.2 below).

\subsubsection {Variations in the UV(IUE) spectra of IRAS16206-5956 (SAO 243756) and
IRAS 18062+2410 (SAO 85766)}

IRAS 16206-5956 was found to be variable in the UV (Fig. 1). The 
spectrum of the star has changed from 21 July 1988 to 12 March 1994 
and from 12 March 1994 to 28 April 1994 suggesting both long term
and short term variability. The maximum flux in the UV was observed
on 12 March 1994. Schild et al (1983) found it to be variable in the optical with 
$\Delta$V=0.13. Since the V magnitudes of the star at the epochs 
of the UV(IUE) observations are not known, we used a mean V magnitude 
(=9.76) from the photometric and spectroscopic database for Stephenson-Sanduleak 
Luminous Stars in the Southern Milky Way (Reed, 1998). 
Its spectral type in literature is listed as A3Iabe (Humphreys, 1975, 
Parthasarathy et al., 2000a) and A0Iae (Schild et al. 1983, Garrison et al., 1977).
Oudmaijer (1996) listed it as B8Ia, citing the Simbad database. However,
we could not find a reference for the same. We compared the dereddened IUE spectra of 
the star at different epochs with the dereddened spectra of 
A3Ib (HD104035) and A0Ia (HD21389) standard stars (Fig. 5a). A match could not 
be obtained in either case. Finally, we adopted the A0Ia standard star (HD21389) 
and modelled the circumstellar extinction in the case of SAO 243756 (as outlined in Sec. 3.2.1)
for the spectra taken on 12 March 1994 (maximum observed IUE flux) and 21 July 
1988 (minimum observed IUE flux). The circumstellar extinction was found to be 
linear in $\lambda^{-1}$ at both epochs (Fig. 5b, Table 4b).

\begin{figure*}
\epsfig{figure=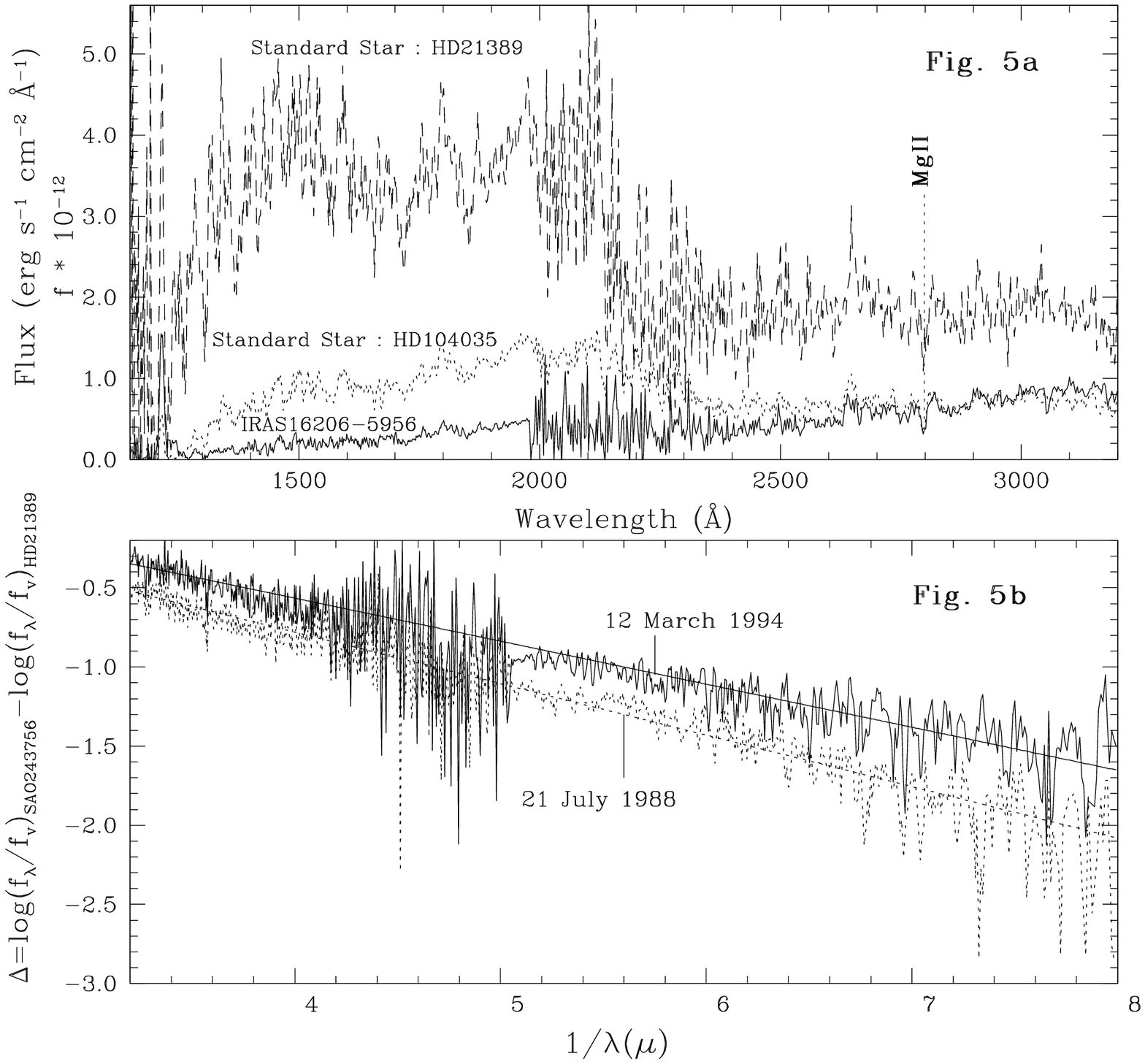,width=17cm,height=20cm}
\caption{IRAS16206-5956 (SAO243756) and the standard A3Ib (HD104035)
and A0Ia (HD166937) stars were dereddened using the 2200\AA~ feature 
in the UV. Fig. 5a shows the dereddened 12 March 1994 spectrum of 
SAO 243756 (solid line) alongwith the dereddened spectra of the 
standard A0Ia (dash line) and A3Ib stars (dotted line).
SAO 243756 is fainter in the spectrum taken on 21 July 1988 (ref. Fig. 1). 
Adopting the A0Ia standard star, Fig. 5b shows the modelled circumstellar 
extinction for the spectrum on 12 March 1994 (solid line) and 21 
July 1988 (dotted line) respectively}
\end{figure*}

The UV flux from IRAS18062+2410 has decreased in Sept. 1995, compared
to the flux from the hot central star in April, 1992 (Fig. 1).
From an analysis of the high resolution optical spectra of the
star, Parthasarathy et al. (2000b) and Mooney et al. (2002), found it to be
metal-poor ([M/H] $\sim$ $-$0.6). Since there are no hot (OB-spectral types), metal-poor
standard stars in the UV, we compared the UV spectrum of this star with 
a Kurucz model closest to the effective temperature, gravity and metallicity
of the star in the optical (Parthasarathy et al. 2000b, Mooney et al., 2002)
We adopted T$_{\rm eff}$=23000K, log g=3.0 and [M/H]=$-$0.5 (Fig. 6a).
Arkhipova et al. (1999, 2000) found that it shows irregular rapid light
variations in the optical with an amplitude of upto 0\fm3. in V. We adopted 
mean V = 11\fm54 (Arkhipova et al., 1999). 
Following the prescription in Sec. 3.2.1, we find that the circumstellar
extinction varies as $\lambda^{-1}$ (Fig. 6b, Table 4b) and has increased
in magnitude from 1992 to 1995.
 
\begin{figure*}
\epsfig{figure=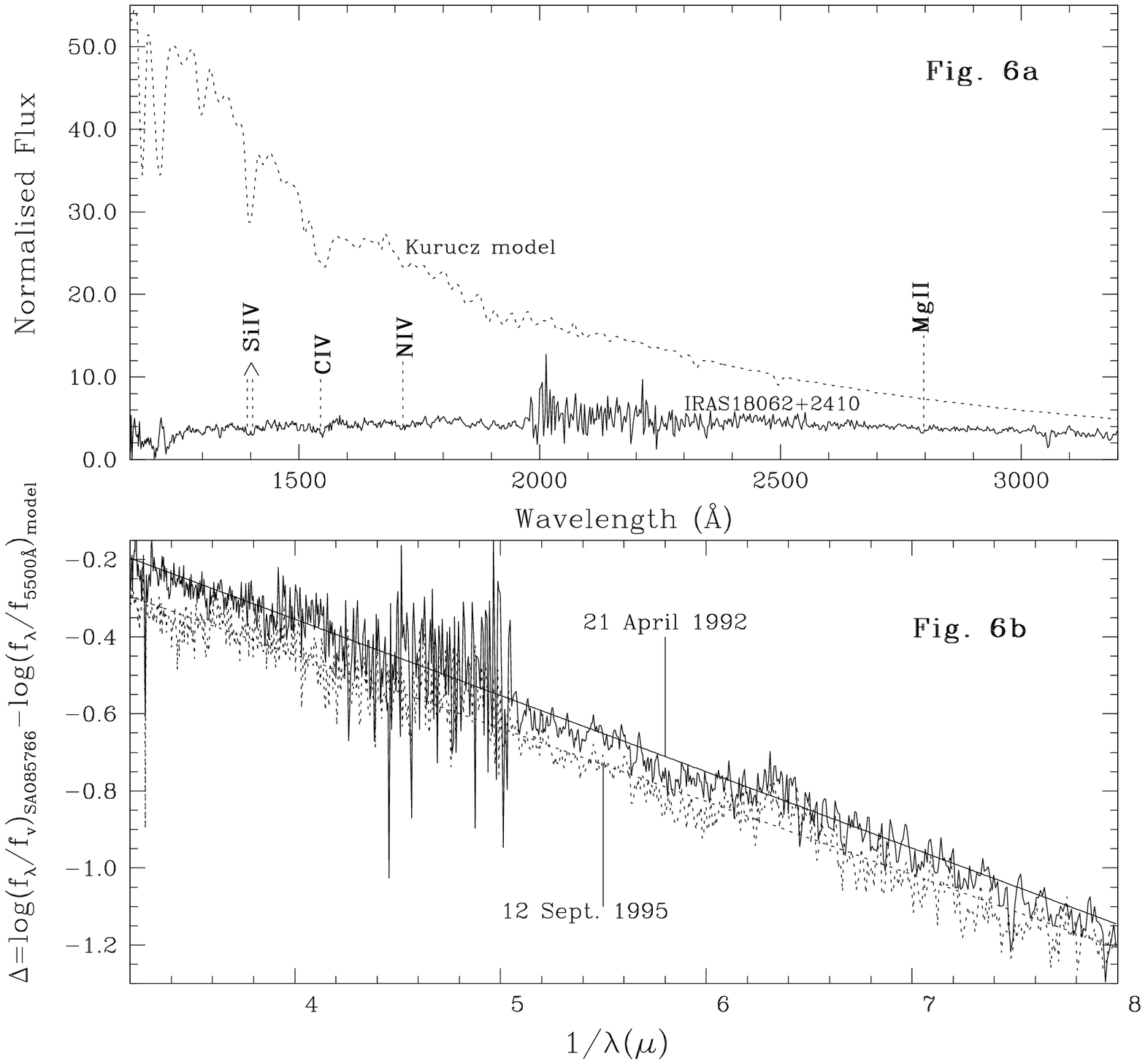,width=17cm,height=20cm}
\caption{IRAS18062+2410 (SAO 85766) was dereddened using the 2200\AA~ feature in the UV. 
Fig. 6a shows the dereddened 1992 spectrum of SAO 85766 normalised to its V-band 
flux (solid line) alongwith Kurucz model (dotted line) of T$_{\rm eff}$=23000K, 
log g=3.0, [M/H]=$-$0.5 normalised to the flux at 5500\AA~. 
SAO 85766 has become fainter in the 1995 spectrum of the
star. Fig. 6b shows the modelled circumstellar extinction for 1992 (solid line)
and 1995 (dotted line).}
\end{figure*}

\subsection{Central star parameters and Energy Budget}

For the 3 hot post-AGB candidates with negligible circumstellar extinction,
IRAS17203-1534, IRAS17460-3114 (SAO 209306) and IRAS18379-1707 (LSS 5112),
we modelled the spectra using solar metallicity Kurucz (1994) model 
atmospheres (Fig. 7) and derived the effective temperatures and gravities 
of these stars. The V-band fluxes of the stars were corrected for extinction 
assuming the normal interstellar extinction law, i.e. 
A$_{v}$ = 3.1 $\times$ E(B$-$V)$_{\rm 2200\AA~}$ (see eg. Seaton, 1979). 
The stellar flux distributions normalised to the corrected V-band flux of each star were compared
with Kurucz models normalised to the respective model's flux at 5500\AA~.
Log g in the case of the O8III star, IRAS17460-3114 (SAO 209206) was estimated to be 4.0. This
value is uncertain due to the non-availability of Kurucz models of lower gravity
at the high temperature (T$_{\rm eff}$=35000K) of the star. The star may also be
slightly metal deficient as discussed in Sec. 3. For the 
remaining stars, we obtained the effective temperatures and gravities based
on their optical spectral types (Lang, 1992). 
Table 5 lists the adopted T$_{\rm eff}$ and log g values. The T$_{\rm eff}$ and log g
values estimated from the optical spectral types of the stars may be uncertain
by $\sim$ $\pm$ 1000K and $\pm$ 1.0 respectively.  High resolution optical 
spectra of these stars are required for an accurate determination of 
T$_{\rm eff}$ and log g. The stars were placed on Sch\"onberber's
(1983, 1987) post-AGB evolutionary tracks for core masses (M$_{c}$) of
0.546, 0.565, 0.598 and 0.644 M$_{\odot}$ (Fig. 8).

The far-IR flux distributions of the hot post-AGB candidates must necessarily
be due to flux from the hot central stars absorbed and re-radiated by the
cold circumstellar dust envelopes which are a remnant of mass-loss on the 
AGB phase of the star. Hence, the integrated far-IR flux (F$_{\rm fir}$) must
be comparable to or less than the integrated stellar flux (F$_{\rm star}$). 
We estimated F$_{\rm fir}$ from the 12$\mu$ to 100$\mu$ IRAS flux distributions 
for the stars. The integrated stellar flux (F$_{\rm star}$) from 1150\AA~ to 5500\AA~ 
was estimated by combining the IUE spectra with the U,B,V magnitudes of the 
stars from literature. The IUE spectra and the U,B,V magnitudes were 
corrected for interstellar extinction derived from the 2200\AA~ feature. 
IRAS17423-1755 was not detected in the UV. Its U,B,V,R,I
magnitudes (Gauba et al., 2003) were corrected for interstellar 
extinction using the standard extinction law (Rieke \& Lebofsky, 1985) 
and the integrated stellar flux from 3650\AA~ (U-band) to 9000\AA~ (I-band)
was estimated. In the three cases with negligible circumstellar extinction 
(IRAS 17203-1534, IRAS17460-3114 and IRAS18379-1707), 
F$_{\rm fir}$ was comparable to or significantly less than F$_{\rm star}$. In
contrast stars with circumstellar extinction in the UV,
eg. IRAS14331-6435, IRAS17074-1845, IRAS17311-4924, IRAS17423-1755, IRAS18062+2410 and 
IRAS22023+5249) showed a high ratio of F$_{\rm fir}$/F$_{\rm star}$ indicating partial 
obscuration of the central stars. These stars may have dusty circumstellar disks.

\begin{figure*}
\epsfig{figure=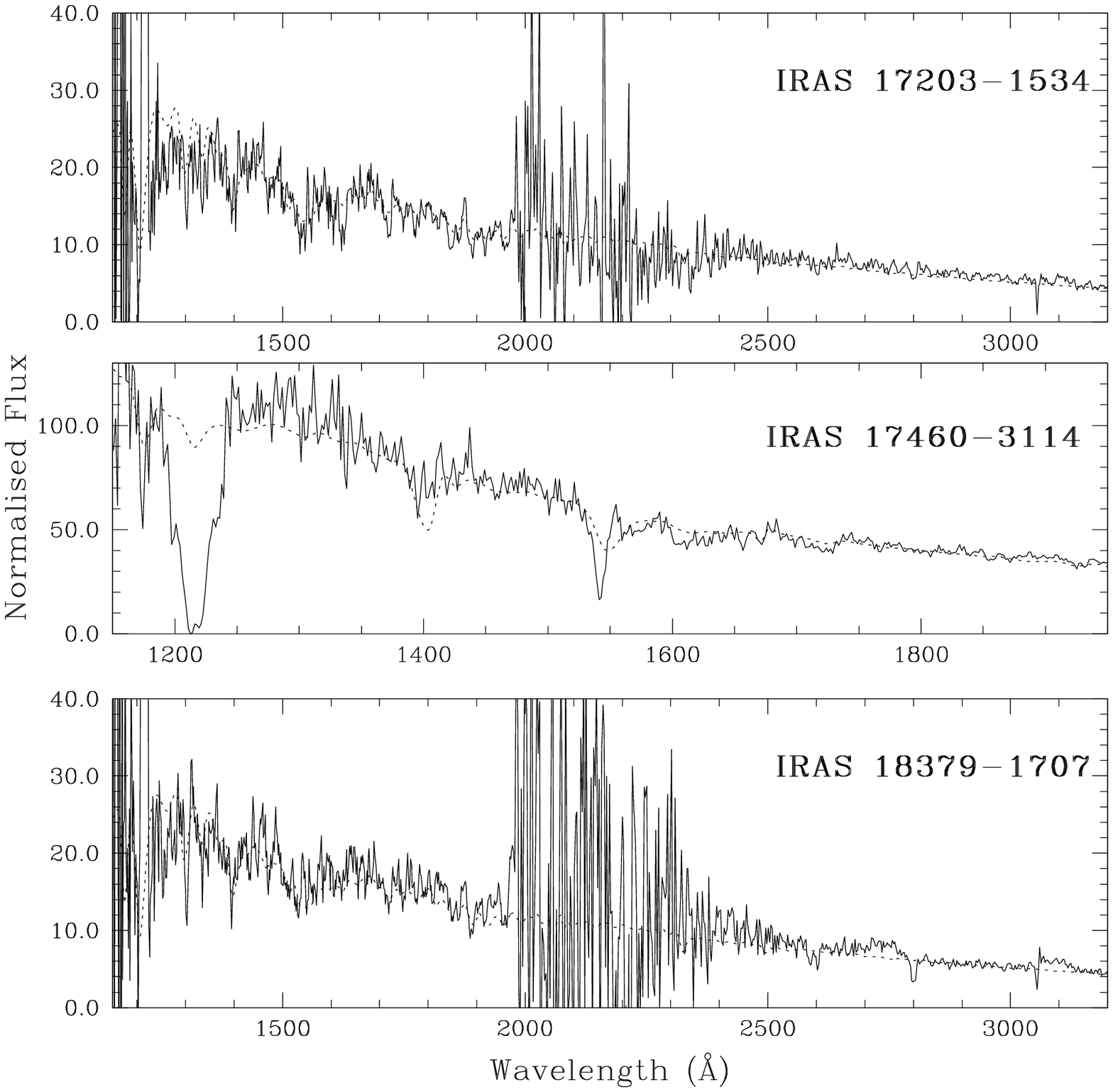,width=17cm,height=20cm}
\caption{The dereddened UV(IUE) spectra of hot post-AGB candidates 
with negligible circumstellar extinction (solid line), normalised to their V-band fluxes are 
plotted alongwith solar metallicity Kurucz (1994) model spectra (dotted line), normalised
to each model's flux at 5500\AA~.}
\end{figure*}

\begin{figure*}
\epsfig{figure=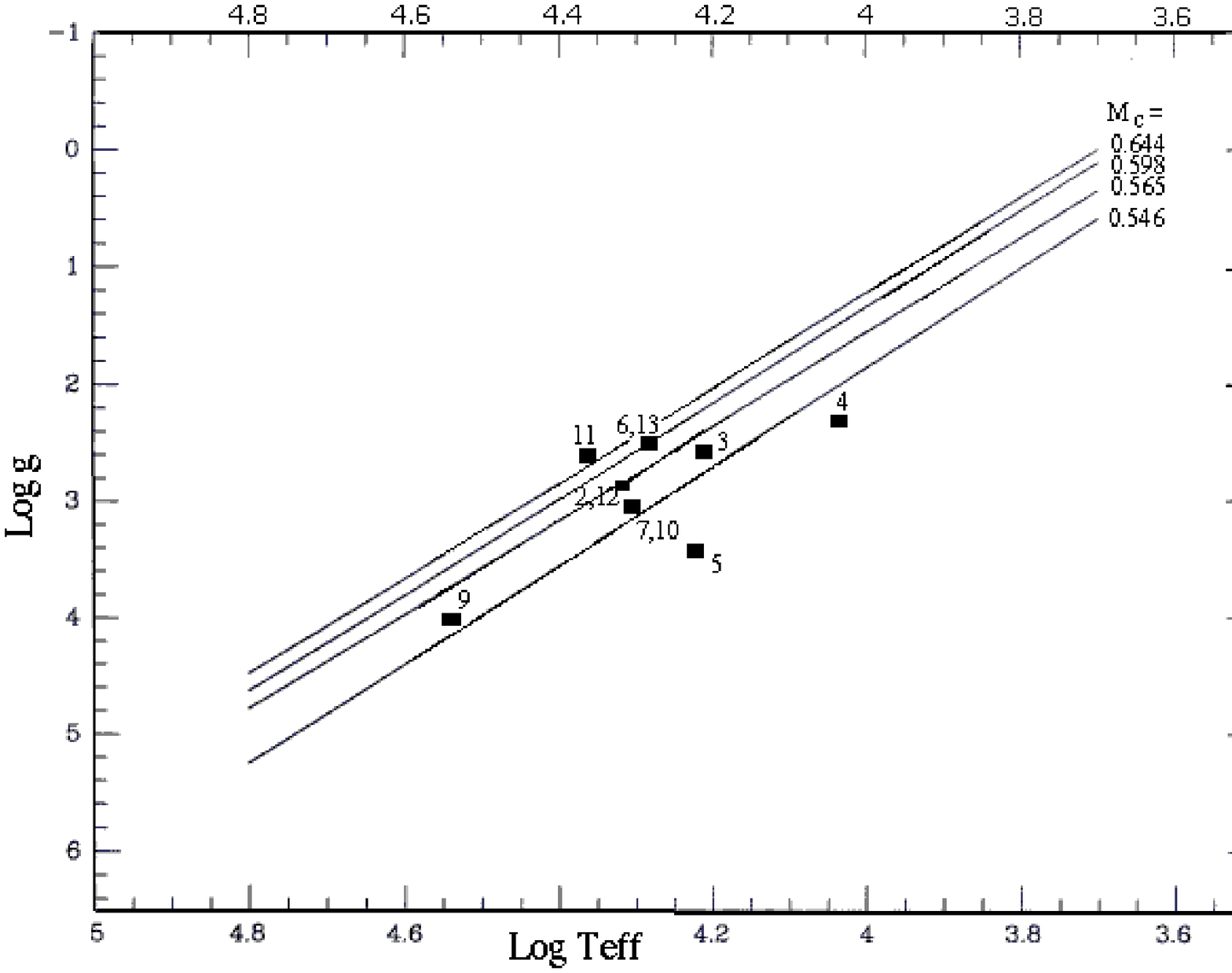,width=17cm,height=13cm}
\caption{Positions of the hot post-AGB candidates on the 
Log g$-$Log T$_{\rm eff}$ 
diagram showing the post-AGB evolutionary tracks 
of Sch\"onberner (1983, 1987) for core masses of 0.546, 0.565, 0.598 and 
0.644M$_{\odot}$ }
\end{figure*}

\begin{table*}
\begin{center}
\caption{Central star parameters and Energy budget}
\begin{tabular}{|c|c|c|c|c|c|c|c|} \hline
Star No. & IRAS & T$_{\rm eff}$(K) & log g & F$_{\rm fir}$ & F$_{\rm star}$ & F$_{\rm fir}$/F$_{\rm star}$ & Wind velocity \\ 
         &      &              &       & X10$^{-12}$ Wm$^{-2}$ & X10$^{-12}$ Wm$^{-2}$ & & kms$^{-1}$ \\
\hline \hline

1. & 12584-4837       & --            & --          & 10.90 & 6.07 & 1.79 & -3769\\
2. & 13266-5551       & 20800         & 2.8         &  5.41 & 9.29 & 0.58 & -1821\\
3. & 14331-6435       & 16200         & 2.6         & 15.40 & 4.50 & 3.42 & --\\
4. & 16206-5956       & 11200         & 2.3         &  1.73 & 2.61 (21/07/88) & 0.66 & --\\
   &                  &               &             &       & 2.97 (12/03/94) & 0.58 & --\\  
5. & 17074-1845       & 17100         & 3.4         & 1.67  & 1.44 & 1.16 & --\\ 
6. & 17203-1534$^{1}$ & 19000$\pm$1000& 2.5$\pm$0.5 &  1.51 & 8.86 & 0.17 & -2402\\
7. & 17311-4924       & 20300         & 3.0         & 21.80 & 4.01 & 5.44 & -1066\\  
8. & 17423-1755       & --            & --          &  6.41 & 1.17 & 5.48 & --\\
9. & 17460-3114$^{1}$ & 35000$\pm$2500& 4.0$\pm$1.0 &  4.95 & 876.8& 0.006 & -1463\\
10.& 18023-3409       & 20300         & 3.0         &  0.58 & 1.13 & 0.51 & -2048\\
11.& 18062+2410$^{2}$ & 23000         & 2.6         &  2.90 & 1.64 (21/04/92) & 1.77 & -- \\
   &                  &               &             &       & 1.33 (12/09/95) & 2.18 & -- \\
12.& 18371-3159       & 20800         & 2.9         &  0.93 & 2.17 & 0.43 & --\\
13.& 18379-1707$^{1}$ & 19000$\pm$1000& 2.5$\pm$0.5 &  3.20 & 16.96& 0.19 & --\\  
14.& 22023+5249       & --            & --          &  3.45 & 1.14 & 3.03 & -3978\\
15.& 22495+5134       & --            & --          &  1.74 & 3.72 & 0.47 & --\\ 

\hline
\end{tabular}

\vspace{0.2cm}

\noindent \parbox{16cm}{$^{1}$ The effective temperatures and gravities 
were estimated using solar metallicity Kurucz (1994) models.\\
$^{2}$ Effective temperature and gravity are from Mooney et al. (2002)}

\end{center}
\end{table*}

\subsection{Terminal wind velocity}

The wavelength at which the shortward edge of the CIV (or NV) absorption
profile intersects the stellar continuum is usually used to estimate the terminal
wind velocities (v$_{\infty}$) of hot stars from low resolution IUE spectra. 
However, this value is usually an upper limit to the true v$_{\infty}$. Alternatively,
the difference between the absorption and emission line centers ($\Delta$v) are 
used to estimate v$_{\infty}$ from low resolution IUE spectra (Prinja, 1994).  
But high dispersion studies (see eg. Perinotto et al., 1982) have shown
that $\Delta$v is less than v$_{\infty}$.  

Considerable circumstellar extinction in 10 of the 15 stars discussed in this paper, 
has significantly distorted the continuum flux distributions and line intensities 
of the CIV profiles. Hence, the point where the blue absorption edge of the 
CIV profile intersects the UV continuum is not reliable. These factors also contribute 
to the fact that the CIV P-Cygni profiles are not clearly observed in the low resolution 
spectra of these stars. The emission peaks may be too weak or lost in the 
reddened continuum. Hence it is not possible to estimate v$_{\infty}$
from the shortward edge of the CIV absorption profile or from the difference
between the absorption and emission line centers.

The CIV resonance doublet (1548\AA~, 1550\AA~) is not resolved in the
IUE low resolution spectra. Stellar wind may be assumed to be absent
in a B0V star. We measured the absorption minimum of the CIV feature
in the UV(IUE) spectrum of a B0V standard star, HD36512 (1548.4\AA~)
from the atlas by Heck et al. (1984). The difference between this
wavelength and the absorption minima of the CIV feature in the 
IUE spectra of our hot post-AGB candidates was used to 
estimate the wind velocities in these stars. If instead, we had used
the mean laboratory wavelength of the CIV resonance doublet (1549\AA~),
the estimated terminal wind velocites would have been greater by 116kms$^{-1}$.
The NV resonance doublet (1238\AA~, 1242\AA~) is also unresolved in low resolution
IUE spectra. Moreover, the NV feature is often contaminated by 
geocoronal Lyman $\alpha$. From the standard star atlas (Heck et al., 1984)
we were unable to find a star in which the NV line is distinguishable 
from Lyman $\alpha$ and unaffected by stellar wind. Hence, in the case
of IRAS12584-4837, we estimated the wind velocity from the difference
between the NV absorption minimum and the mean laboratory wavelength
of the line (1240\AA~). The estimated wind velocities are listed in Table 5. 
No shift was observed in the case of IRAS17074-1845, IRAS18062+2410, IRAS18379-1707 and
IRAS22495+5134. CIV (or NV) lines were not observed in IRAS14331-6435 and 
IRAS16206-5958. The SiIV (1394\AA~) absorption line in IRAS14331-6435 did not show 
a wavelength shift due to stellar wind. For low resolution IUE spectra, the FWHM 
of the instrumental width is 5\AA~ (Castella \& Barbero, 1983). Assuming a 
3\AA~ measurement error in estimating the blue shift of the lines from our
low resolution spectra, would correspond to an error of $\sim$ 580 kms$^{-1}$
and 726kms$^{-1}$ in the estimate of wind velocites from CIV and NV respectively.
However, our terminal wind velocities exceeded this error estimate. So, we may 
conclude that although a crude estimate is obtained of the terminal wind velocities, 
it nevertheless indicates the presence of stellar winds and post-AGB mass-loss
in some of these stars.

\section{Notes on individual objects :}

\#IRAS 12584-4837 (Hen3-847)\\
Henize (1976) identified it as an H$\alpha$ emission line star.
Based on its far infrared flux distribution and high galactic
latitude (b=$+$13.95$^{\circ}$) Parthasarathy(1993a) classified
it as a post-AGB star. It was found to be variable in the optical
(Kazarovets et al., 2000, de Winter et al., 2001). The Hipparchos magnitudes at 
maximum and minimum are 10\fm52 and 10\fm70 respectively.  
Th\'e et al. (1994) listed it as a candidate Herbig Ae/Be star. 
The presence of HeII line (1640\AA~) indicates a very high
temperature. The UV spectrum and the presence of several emission lines 
(Fig. 2) suggest that it may be a massive young star. Another possibility is
that it may be a luminous blue variable. Photometric monitoring and high resolution
spectroscopy may enable us to further understand the evolutionary stage of
this star. From the NV profile we estimated a terminal wind velocity 
of $-$3769 kms$^{-1}$.

\#IRAS 13266-5551 (CPD-55 5588)\\
It was classified as a post-AGB star on the basis of its far infrared flux 
distribution (Parthasarathy, 1993a). From the optical spectra, Parthasarathy
et al. (2000a) classified it as B1Ibe. They found the Balmer lines in emission
upto H$\delta$. The LWP spectrum of the star is saturated from
2600\AA~ to 3000\AA~. The central star has a terminal wind velocity of 
$-$1821 kms$^{-1}$. 

\#IRAS 14331-6435 (Hen3-1013)\\
Henize(1976) identified it as an H$\alpha$ emission line star. It is listed
as a Be star in Wackerling (1970). Based on its far infrared flux 
distribution it was classified as a post-AGB star (Parthasarathy \& Pottasch 1989, 
Parthasarathy 1993a). Assuming a Be spectral type, Kozok (1985b) derived a distance of 
1.9 kpc to the star.  Loup et al. (1990) detected CO emission in this object and
derived an expansion velocity of 15kms$^{-1}$. They derived a distance of 1.3kpc
using the total IR flux in IRAS bands and assuming a post-AGB luminosity of 
10$^{4}$L$_{\odot}$. Parthasarathy et al. (2000a) found it to be a 
B3 supergiant (B3Ie) with H$\beta$ in emission and H$\gamma$ filled in. 
Modelling yielded a circumstellar extinction value
of 0\fm26. The F$_{fir}$/F$_{star}$ (= 3.42) ratio also suggests
obscuration of the hot central star.  

\#IRAS 16206-5956 (SAO 243756)\\
On the basis of IRAS data, high galactic latitude, and supergiant spectrum,
Parthasarathy (1993a) considered it to be a post-AGB star.  
H$\alpha$ and H$\beta$ were detected in emission (Oudmaijer 1996, 
Parthasarathy et al., 2000a). The variation in the UV
may be attributed to variable circumstellar extinction. 

\#IRAS 17074-1845 (= Hen3-1347)\\
Henize (1976) identified it as an H$\alpha$ emission line object.
It was classified as a hot post-AGB star on the basis of its high
galactic latitude, far-IR colors similar to PNe and Be spectral
type (Parthasarathy, 1993a). It was classified as a B3IIIe star
(Parthasarathy et al., 2000a). 

\#IRAS 17203-1534\\ 
On the basis of its high galactic latitude, early spectral type (B1IIIpe),
Balmer line emission (H$\beta$) and far-IR colors similar to PNe, Parthasarathy
et al. (2000a) classified it as a possible post-AGB star. We found that the
UV spectrum closely follows that of a standard B1II star (HD173502). Using
solar metallicity Kurucz (1994) models we obtained T$_{\rm eff}$ = 19000 $\pm$ 1000K
and log g = 2.5 $\pm $0.5. 

\#IRAS 17311-4924 (Hen3-1428)\\
It was classified as a post-AGB star on the basis of its IRAS data 
(Parthasarathy \& Pottasch 1989, Parthasarathy 1993a). Loup et al.
(1990) detected CO emission in this object typical for circumstellar 
shells around evolved objects. They derived a distance of 1.1 kpc and an 
expansion velocity of 11 kms$^{-1}$. Nyman et al. (1992) found
an expansion velocity of 14.1 kms$^{-1}$ from CO observations similar to that
found by Loup et al. (1993). Assuming a Be star, Kozok (1985b) 
derived a photometric distance of 2.6kpc. Parthasarathy et al. (2000a)
classified it as B1IIe. We found no apparent change between the
UV(IUE) spectra taken in 1988 and 1993. 

\#IRAS17423-1755 (= Hen3-1475)\\
It was found to be a B-type emission line star (Henize, 1976). On the basis 
of IRAS data, Parthasarathy \& Pottasch (1989) first classified
it as a hot post-AGB star. It is a bipolar PPN
(Bobrowsky et al. 1995, Riera et al. 1995, Borkowski et al., 1997) with 
wind velocities greater than 1000 km~s$^{-1}$ (S\'anchez Contreras \& Sahai, 2001,
Borkowski \& Harrington, 2001). Borkowski et al. (1997) concluded that Hen3-1475 
is a point symmetric nebula. HST images of the object revealed 
outflows perpendicular to the dusty torus obscuring the hot central star.
From the spectral energy distribution, Gauba et al. (2003) 
detected a hot dust component ($\sim$ 1000K) indicating cirumstellar dust close 
to the central star as a result of ongoing post-AGB mass-loss. The non detection
of the B-type central star in the SWP spectrum (35 minutes exposure) may be
due to obscuration of the star by the dusty torus seen in the HST image. 
The F$_{\rm fir}$/F$_{\rm star}$ (= 5.48) ratio also suggests obscuration of the hot
central star by the dusty disk.

\#IRAS 17460-3114 (SAO 209306)\\
It shows far-IR colors similar to PNe (Parthasarathy, 1993a). Based on low
resolution spectra, it was classified as O8III (Parthasarathy et al., 2000a). 
In the UV(IUE) spectrum of the star, CIV (1550\AA~) shows a P Cygni profile with a 
strong absorption and weak emission component. The NV (1240\AA~) feature 
is contaminated by Lyman $\alpha$. 

\#IRAS 18023-3409 (LSS 4634)\\
It is listed in the LSS (Luminous stars in the southern Milky Way)  
catalogue (Stephenson \& Sanduleak, 1971) as an OB+ star. 
Parthasarathy et al.(2000a) found H$\beta$ in emission and H$\gamma$ 
filled in. They classified it as B2IIIe. The UV(IUE) spectrum of the star
shows considerable circumstellar extinction (E(B$-$V)$_{\rm C.S.}$=0.43). 

\#IRAS 18062+2410 (SAO 85766)\\
Stephenson(1986) identified it as an H$\alpha$ emission line star.
The star appears to have rapidly evolved in the last 20-30 years 
(Arkhipova et al. 1999, Parthasarathy et al. 2000b). It was classified 
as a high galactic latitude, post-AGB star having far-IR colors similar 
to PNe (Volk and Kwok, 1989, Parthasarathy, 1993a). 
According to the HDE Catalog, its spectral type in 1940 was A5.
The UV colors of this star listed in the Catalogue of Stellar 
Ultraviolet Fluxes (Thompson et al., 1978) and based
on observations with the Ultraviolet Sky Survey Telescope (S2/68)
onboard the ESRO satellite TD-1 indicated that in 1973 its spectral type
was still A5I (Parthasarathy et al., 2000b). 
However, the spectral energy distribution of this star
obtained in 1985-87 indicated a Be spectral type. 
(Downes and Keyes, 1988). An analysis of its high resolution spectra,
revealed the underabundance of carbon and metals, high radial velocity
and the presence of low excitation nebular emission lines
(Arkhipova et al. 1999, Parthasarathy et al., 2000b).
On the post-AGB evolutionary tracks of Sch\"onberner (1983, 1987), 
it appears to be evolving into the PN phase with M$_{c}$ $\simeq$ 0.644M$_{\odot}$. 
Variable circumstellar extinction, which may be due to a dusty torus 
in motion around the hot central star may explain the observed variations
in the UV.

\#IRAS 18371-3159 (LSE 63)\\
It is listed as OB+ in the extension to the Case-Hamburg OB-star
surveys (Drilling \& Bergeron 1995). Parthasarathy et al. (2000)
classified it as B1Iabe. Preite-Martinez (1988) classified it as 
a possible new PN and estimated a distance of 2.4 kpc to the star.
On the post-AGB evolutionary tracks of Sch\"onberner (Fig. 8),
it appears to be evolving into the PN stage with
M$_{c}$ = 0.565M$_{\odot}$.

\#IRAS 18379-1707 (LSS 5112)\\
Parthasarathy et al. (2000) classified it as B1IIIpe. The photometry of the
star is from Reed (1998). IRAS colors similar to PNe, high galactic 
latitude, early B-type giant spectra in the UV and optical and
the presence of CII (1335\AA~), SiIV (1394,1403\AA~), CIV(1550\AA~), 
NIV (1718\AA~) and MgII (2800) lines in the IUE spectrum support its 
classification as a hot post-AGB star. Using solar metallicity Kurucz (1994)
models we determined T$_{\rm eff}$=19000 $\pm$ 1000K and log g = 2.5 $\pm$ 0.5.

\#IRAS 22023+5249 (LSIII +5224)\\ 
Its IRAS colors were found to be similar to PNe and 
Parthasarathy et al. (2001) classified it as a hot post-AGB
star. The photometry of the star was obtained from the 
Tycho-2 Catalogue (Hog et al., 2000). 
It is listed in Wackerling's (1970) catalog of early-type emission-line 
stars. The B spectral type of the star was obtained from the Simbad 
database. We compared the UV spectrum of this star with that of a B2I
standard star (HD41117) and found it to be similar. The 
F$_{\rm fir}$/F$_{\rm star}$ (=3.03) ratio suggests
obscuration of the hot central star and lends support to the modelled 
circumstellar extinction value of 0\fm44.  There was
no difference between the 1993 and 1995 UV(IUE) spectra.

\#IRAS 22495+5134 (LSIII +5142)\\
It is classified as a PN in the Strasbourg-ESO catalogue of Galactic 
planetary nebulae (Acker et al., 1992). Tylenda and Stasinska (1994) 
reported an angular diameter of 0.5\arcsec, expansion velocity 
of 10 kms$^{-1}$ and V=12.08. The V magnitude from the Tycho-2 Catalogue 
(Hog et al., 2000) is 11\fm78. Handler (1999) found it to be variable 
with amplitude variations of 0\fm3 in the Johnson V band. While the 
long term variations (several days) were non periodic, the short
term variations were quasi-periodic with time scales of either
8.9 or 14.3 hours. Variations in the stellar mass-loss coupled with
stellar pulsation may explain the observed long and short-term variability. 
From the Hipparchos catalog, the Hipparchos magnitudes at maximum
and minimum are 12\fm11 and 12\fm47 respectively.
From the logarithmic extinction at H$\beta$ using radio flux at 5Gz (c=0.41), 
Kaler(1983) obtained  E(B-V)=0\fm28. Using the H$\alpha$ to 
H$\beta$ ratio (c=0.55), Tylenda et al. (1992) obtained E(B-V)=0.37. 
These values are in good agreement with the interstellar E(B-V)=0.33 
derived from the 2200\AA~ feature in the UV. 

\section{Discussion \& Conclusions}

We analysed the UV(IUE) spectra of 15 hot post-AGB candidates. In 11
cases (IRAS13266-5551 (CPD-55 5588), IRAS14331-6435 (Hen3-1013), IRAS16206-5956 (SAO 243756),
IRAS17074-1845 (Hen3-1347), IRAS17311-4924 (Hen3-1428), IRAS17423-1755 (Hen3-1475), 
IRAS18023-3409 (LSS 4634), IRAS18062+2410 (SAO 85766), 
IRAS18371-3159 (LSE 63), IRAS22023+5249 (LSIII +5224) and IRAS22495+5134 (LSIII +5142)), 
the UV spectra revealed obscuration of the hot central stars due to
circumstellar dust. While IRAS17423-1755 (Hen3-1475) was not detected at all in a 
35 minute exposure, the UV continua of the remaining 10 stars were found 
to be considerably reddened. We found that the circumstellar extinction
in these 10 stars varies linearly as $\lambda^{-1}$. A $\lambda^{-1}$
law for the circumstellar extinction was also found in the case
of the post-AGB star, HR4049 (Waters et al., 1989, Monier \& Parthasarathy,
1999). In the context of Mie scattering (Spitzer, 1978), linear extinction
arises from dust grains small compared to the wavelength of light. 
The shortest wavelength of light at which the extinction is linear
can give an estimate of the size of the smallest grains in the circumstellar
environment of these stars ($\lambda$ = 2$\pi$a, where, a is the radius 
of the dust grain). However our IUE SWP observations are limited
to 1150\AA~ ($\lambda^{-1}$ $\approx$ 8.7$\mu^{-1}$). Shortward of 1300\AA~
the spectra are noisy and often contaminated by Lyman $\alpha$. Taking
1300\AA~ as the shortest observed wavelength at which the
extinction is linear in $\lambda^{-1}$, we may infer an upper limit 
of ~a $\approx$ 200\AA~ for the radii of the small grains. 
Waters et al. (1989) speculate that the destruction of these  
grains in the vicinity of the hot central stars of PPNe and PNe may 
give rise to smaller grains and polyaromatic hydrocarbons (PAHs).
PAH features at 8.2, 8.6 and 11.3 $\mu$ have been detected in the
circumstellar environment of several post-AGB stars, PPNe and PNe
(see eg. Beintema et al., 1996). It would be interesting to study the 
infrared spectra of our hot post-AGB candidates to know more about the 
chemical compositions (carbon-rich or oxygen-rich nature) of the dust 
grains and the evolution of these grains in the circumstellar environment 
of these stars. Variation of IRAS16206-5956 (SAO 243756)
and IRAS18062+2410 (SAO 85766) in the UV may be due to stellar pulsations
and/or due to variable circumstellar extinction similar to that observed in the case of
HR4049 (Waters et al., 1989, Monier \& Parthasarathy, 1999).
Significant circumstellar extinction was not 
observed in the case of IRAS17203-1534, IRAS17460-3114 (SAO 209306) and 
IRAS18379-1707 (LSS 5112). The effective temperatures and gravities of 
these three stars were estimated using Kurucz model atmospheres. 

F$_{\rm fir}$/F$_{\rm star}$ $>>$ 1.0 in the case of 
IRAS14331-6435 (Hen3-1013), IRAS17311-4924 (Hen3-1428), 
IRAS17423-1755 (Hen3-1475), IRAS18062+2410 (SAO 85766) and IRAS22023+5249 
(LSIII +5224) indicates the presence
of dusty disks around these stars. From the UV(IUE) spectra we found that 
7 (IRAS12584-4837, IRAS13266-5551 (CPD-55 5588), IRAS17203-1534, IRAS17311-4924 (Hen3-1428), 
IRAS17460-3114 (SAO 209306), IRAS18023-3409 (LSS 4634) and IRAS22023+5249 (LSIII +5224)) 
of the 15 hot post-AGB candidates
have stellar wind velocities in excess of 1000kms$^{-1}$ indicating post-AGB mass-loss.

\end{document}